\documentclass[a4paper,11pt]{article}
\usepackage{jcappub} 
\usepackage{comment}
\usepackage{lineno}
\usepackage{subcaption}
\usepackage{orcidlink}

\usepackage{newtxtext,newtxmath}
\usepackage{orcidlink}
\usepackage{braket}

\usepackage[T1]{fontenc}

\usepackage{graphicx}	
\usepackage{amsmath}	

\newcommand*\diff{\mathop{}\!\mathrm{d}}
\newcommand*\Diff[1]{\mathop{}\!\mathrm{d^#1}}

\title{ {\tt SpyDust}: an improved and extended implementation for modeling spinning dust radiation}

\author[1]{Zheng Zhang\note{Corresponding author.}\orcidlink{0000-0002-9154-2803}}
\author{and Jens Chluba}
\affiliation{Jodrell Bank Centre for Astrophysics, University of Manchester, \\
Manchester, M13 9PL, United Kingdom}

\emailAdd{Zheng.Zhang@Manchester.ac.uk}
\emailAdd{Jens.Chluba@Manchester.ac.uk}

\abstract{
This paper presents {\tt SpyDust}, an improved and extended implementation of the spinning dust emission model based on a Fokker-Planck treatment. {\tt SpyDust} serves not only as a {\tt Python} successor to {\tt spdust}, but also incorporates some corrections and extensions. Unlike {\tt spdust}, which is focused on specific grain shapes, {\tt SpyDust} considers a wider range of grain shapes and provides the corresponding grain dynamics, directional radiation field and angular momentum transports. We recognise the unique effects of different grain shapes on emission, in particular the shape-dependent mapping between rotational frequency and spectral frequency. 
In addition, we update the expressions for effects of electrical dipole radiation back-reaction and plasma drag on angular momentum dissipation.
We also discuss the degeneracies in describing the shape of the spectral energy distribution (SED) of spinning dust grains with the interstellar environmental parameters. Using a typical Cold Neutral Medium (CNM) environment as an example, we perform a perturbative analysis of the model parameters, revealing strong positive or negative correlations between them. A principal component analysis (PCA) shows that four dominant modes can linearly capture most of the SED variations, highlighting the degeneracy in the parameter space of the SED shape in the vicinity of the chosen CNM environment. This opens the possibility for future applications of moment expansion methods to reduce the dimensionality of the encountered SED parameter space.
}

\begin{document}
\maketitle
\flushbottom



\newpage
\section{Introduction}
The existence of non-thermal emission from spinning dust grains at high radio frequencies was first proposed in 1957 to explain the observed radio noise from discrete sources \cite{erickson1957}. Since then, spinning dust emission has been studied in several contexts, including environments with conversion of optical photons from stars into radio emission \cite{hoyle1970} and the study of radio-emitting dust in spiral galaxies \cite{ferrara1994}.

The topic received considerable attention with the advent of the era of high-precision cosmological observations. As experiments focused on the Cosmic Microwave Background (CMB) required precise separation of foreground signals from the diffuse microwave sky, \cite{kogut1995high, kogut1996microwave, leitch1997anomalous} discovered a new component of `anomalous' microwave emission (AME), correlated with dust but of uncertain physical origin. This anomaly, together with the known components of the diffuse microwave galactic foreground, prompted further investigation.

Shortly after the discovery of the AME, \cite{DL98a, DL98b} proposed that the AME could be explained by electric dipole radiation from small, spinning dust grains in the interstellar medium (ISM). Their pioneering work provided detailed predictions for spinning dust emission, incorporating a range of angular momentum excitation and damping processes relevant to very small grains. Although their early model did not account for certain physical effects - for instance, the non-spherical grain shapes and internal relaxation - it was in general agreement with observations (see, for example, \cite{lazarian2003microwave}). This breakthrough stimulated further research into the physical mechanisms and spectral characteristics of spinning dust emission.

By applying a Fokker-Planck approach to calculate the angular velocity distribution of dust grains, \cite{AHD09} advanced the treatment of rotational dynamics. This development enabled the calculation of the theoretical spectral energy density (SED) using the publicly available {\tt spdust} code written in Interactive Data Language ({\tt IDL}).
\cite{hoang2010} further refined the rotational dynamics models by incorporating the wobbling motion of disk-like grains, a result of internal fluctuations, and by accounting for transient spin-up due to ion collisions through the use of the Langevin equation. \cite{SAH11} extended {\tt spdust} by modelling the rotation of wobbling, oblate dust grains with random orientations in their updated {\tt spdust2} code.
Henceforth, we will use ``{\tt spdust}'' to refer to both generations of this code and the related literature.
Further developments have included the effects of irregular grain shapes, stochastic heating and emissivity enhancements driven by compressible turbulence \citep{hoang2011}. There have also been studies, notably by \cite{draine1999magnetic, hoang2016unified} and \cite{hensley2017modeling}, of magnetic dipole radiation from ferromagnetic spinning dust, while \cite{draine2016quantum} introduced improvements to account for quantum suppression of dissipation and alignment processes. For a comprehensive overview of the current state of AME research, see \cite{dickinson2018state}.

Thanks to its open-source\footnote{Although the code is open-source, it requires the proprietary {\tt IDL} language for execution.} nature and robust numerical performance - benefiting from the semi-analytic solution of the Fokker-Planck equation - {\tt spdust} has become a widely-used tool for modelling spinning dust emission. However, the increasing complexity of the analysis pipeline and the precision requirements of modern cosmological observations, such as large-scale Bayesian joint analyses for CMB experiments (e.g., \cite{galloway2023beyondplanck}) and studies of CMB spectral distortions (see \cite{kogut2019cmb, Chluba2021Voyage, abitbol_pixie}), have stretched the capabilities of current models. To meet these evolving requirements, a new simulation tool is needed that offers greater code generality, portability (allowing easy replacement or modification of models), and improved numerical efficiency.

In this paper we present a new implementation, {\tt SpyDust},\footnote{Available at \url{https://github.com/SpyDust/SpyDust}.} which is based on {\tt spdust} but has several improvements.
First, the whole package is implemented in {\tt Python}, except for a public {\tt Mathematicamodule } that allows readers to reproduce the formulas in the paper. 
Second, {\tt SpyDust} is a modular package, and users are free to apply their own statistical models, although by default it inherits most of the treatments from {\tt spdust}.
Third, we generalise the treatment of the grain shape to allow for an ensemble of grains with arbitrary oblateness, not just the specific shape as for {\tt spdust}.  
Accordingly, we provide updated expressions for the plasma drag and fluctuation, and the angular momentum dissipation caused by the electrical dipole back-reaction, applicable to different grain oblateness.
Other processes are not directly influenced by $\beta$ but may depend on derived quantities like the volume-equivalent radius a or the effective area radius.
{\tt SpyDust} maps its grain parameters to those in {\tt spdust} under assumed geometries - elliptical cylinders or ellipsoids, which generalize the perfect disc and sphere models of {\tt spdust}.
Calculations for these processes follow the {\tt spdust} framework, with the generalizations ensuring that {\tt spdust} remains a special case of {\tt SpyDust}.

Despite all its advantages, inheriting {\tt spdust} does come with its drawbacks. Due to the limitations of the Fokker-Planck equation, impulsive torques on the grains are not taken into account. This could be problematic for characterising the contribution of the smallest grains to the high frequency end. Also, the internal alignment of the angular momentum is assumed to be isotropic, which is a good assumption for the diffuse ISM phases [Cold Neutral Medium (CNM), Warm Neutral Medium (WNM), Warm Ionized Medium (WIM)] and high radiation intensity phases [Reflection Nebula (RN), Photodissociation Region (PDR)], but not for low radiation density regions [Dark Cloud (DC), Molecular Cloud (MC)]  \cite{SAH11}. 
If these issues are of concern, users are recommended to use their own model-defined distribution functions in {\tt SpyDust}.

In addition, we recognise that the discussion of plasma drag could be further refined. Currently, the dissipation rate for plasma drag is derived using the detailed balance rule, a method whose validity depends on an accurate and comprehensive characterisation of the thermal baths. 
In this context, one can explicitly take into account not only the thermal bath of the ionic system, but also the internal processes that disturb the internal alignment ($m$ substates). However, a more thorough revision would require a detailed Fokker-Planck treatment, taking into account the internal alignment driven towards equilibrium at a finite temperature. We defer this updated treatment to a separate paper. In the present work, we thus adhere to the existing plasma drag framework of {\tt spdust}, but explicitly update the rates to account for grain oblateness.

In contrast to using explicit distribution functions for forward simulation, an alternative approach follows a kind of `backward fitting' logic. In this method, one acknowledges the uncertainties regarding grain and environmental conditions, such as grain size and shape distributions. A perturbative statistical analysis, such as the moment expansion method \citep{Jens17, vacher2023high, carones2024optimization}, is then employed to model the SED using fundamental spectra and moment coefficients. The model is then fit to observational data, allowing the coefficients to be determined and the SED of the spinning dust to be reconstructed.
We present the work following this approach in a separate paper but do highlight some of the future opportunities already here.
On the other hand, the use of derivative spectra to linearly represent the SED space provides an effective way to explore the fundamental modes of SEDs. As a preview of the dedicated follow-up work, we discuss parameter degeneracy in the SED space of spinning dust. Using a simple example, we find that three to four modes are sufficient to capture most of the features of the SED, even though forward modelling requires more than ten parameters.

In this paper, we limit the discussion of previous ingredients and focus on the changes with respect to {\tt spdust}, highlight new methods introduced in {\tt SpyDust} and discuss them where necessary. 
In Section~\ref{sec: dynamics} we formalise the dynamics of arbitrarily shaped dust grains, the electric dipole radiation and its back-reaction, and the general form of the overall SED. In Section~\ref{sec: config distribution} we discuss the distribution of the rotational configuration parameters, in particular the angular momentum transport, using the Fokker-Planck equation. In the same section, we update the expressions for the radiative damping and the plasma drag and fluctuation effects.
In Section~\ref{Sec: SED}, we introduce the formalism for synthesizing the SED and present various analyses to illustrate the impact of the corrections and extensions implemented in {\tt SpyDust}. We also examine the degeneracies within the parameter space in the same section.
In Section~\ref{sec: conclusion} we summarise what has been done in this work and describe the final conclusions.

\section{Radiation from Spinning Dust Grains}
\label{sec: dynamics}
This section begins with the rotational dynamics of irregular dust grains, where we explicitly give the ordinary differential equations (ODEs) describing the time evolution of the Euler angles for torque-free grains.
The rotating system is characterised by angular momentum ($L$), external alignment ($\phi_L, \theta_L$), internal alignment ($\phi_b, \theta_b, \psi_b$), grain size ($I_{\rm ref}$), in-plane ellipticity ($\alpha$) and oblateness ($\beta$). 
These parameters define the arbitrary rotation of any grain in the observer’s reference frame. Although this work does not address the anisotropic external alignment caused by systematic torques, nor does it consider the case of $\alpha\neq 0$, our formulation lays the groundwork for future extensions to these scenarios.
We then discuss the electric dipole radiation and the general form of the overall SED, where we have generalised the grain shape to an ensemble of oblateness. 

\subsection{Grain dynamics and electric dipole emission}
To facilitate the discussion of the dynamics of rotating dust grains and electric dipole radiation, we define three sets of spatial orthogonal basis vectors:
\begin{enumerate}
    \item The observer-preferred basis, denoted as $O_{\rm obs}$. This basis does not change with time and takes the line of sight as the $z$-axis.
    \item The angular momentum based basis, denoted $O_L$, where the $z$-axis is aligned with the direction of the angular momentum $\boldsymbol{L}$. This basis is defined for each grain independently.
    \item The grain body basis, denoted as $O_b$. This is a per-grain defined rotational basis that is static relative to the grain and is used to describe the fundamental properties of the dust grain, such as its moments of inertia and the electric dipole.
    The basis is chosen so that the moments of inertia are diagonal in this basis.
\end{enumerate}
\begin{figure}[ht]
  \subcaptionbox*{$O_L$ basis and $O_b$ basis}[.5\linewidth]{%
    \includegraphics[width=\linewidth]{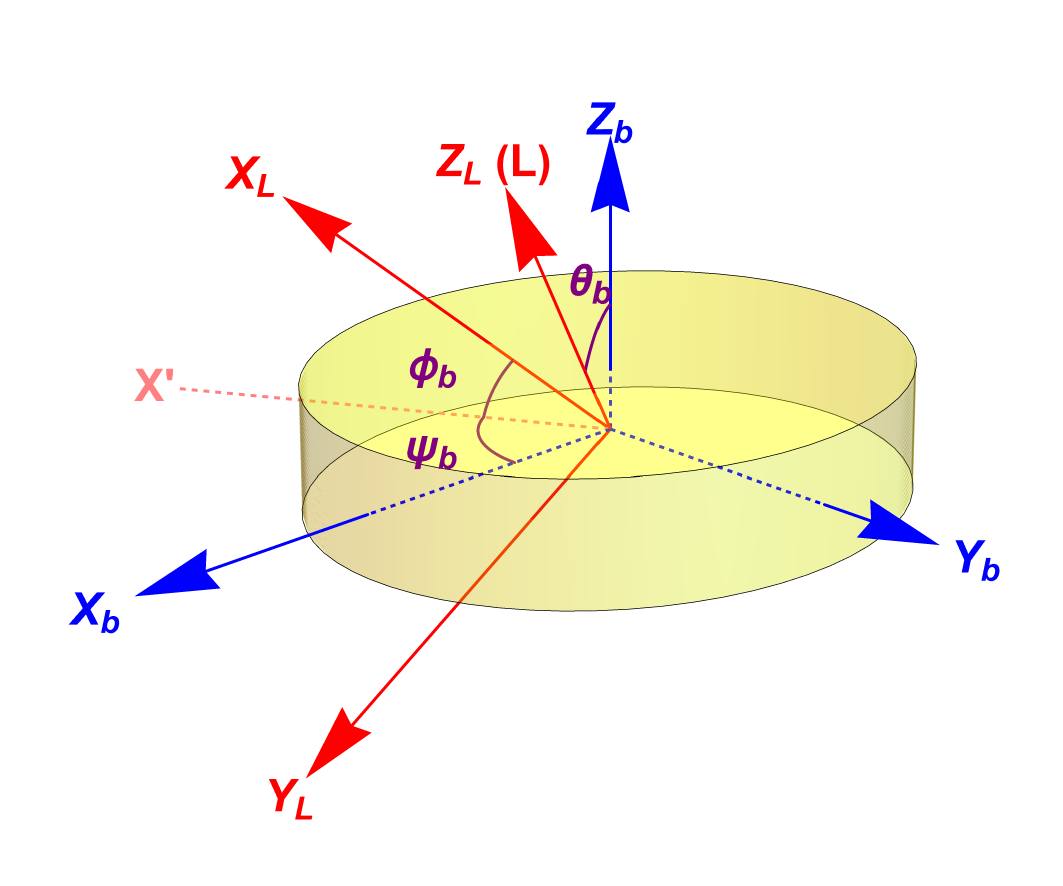}%
  }%
  \hfill
  \subcaptionbox*{$O_L$ basis and $O_{\rm obs}$ basis}[.5\linewidth]{%
    \includegraphics[width=\linewidth]{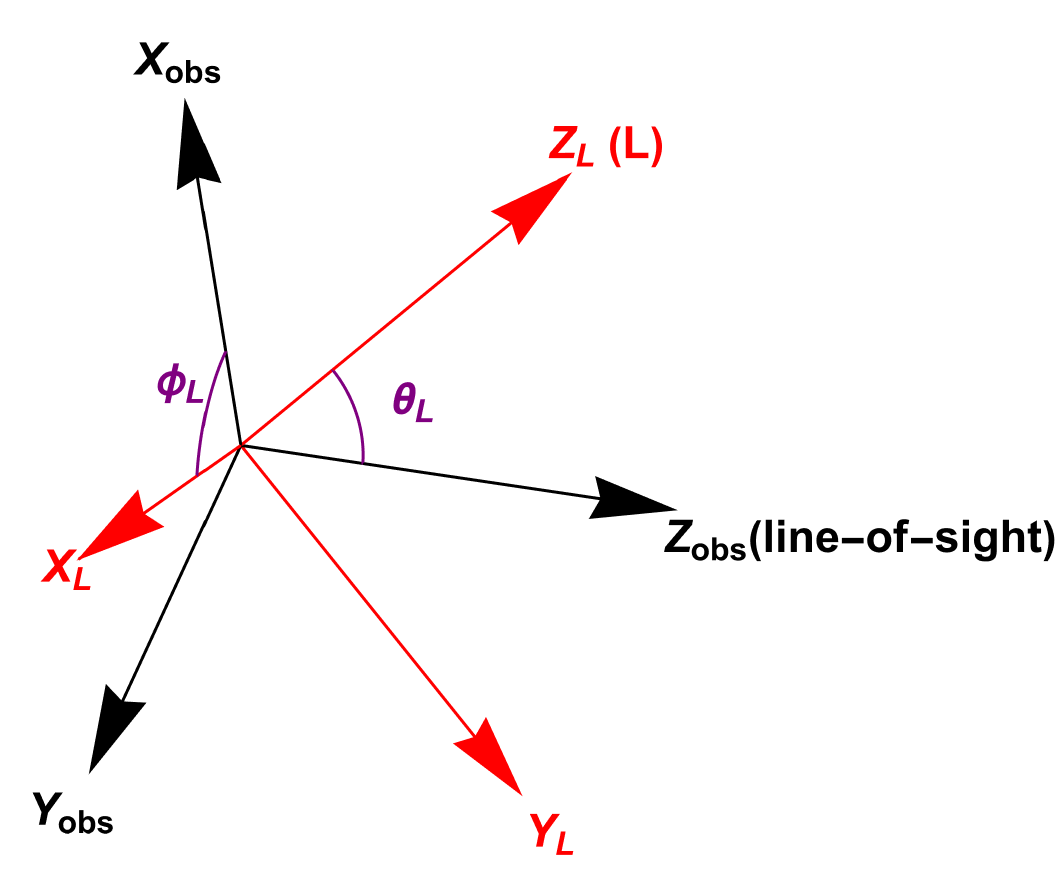}%
  }
  \caption{Euler angle transformations between three coordinate bases relevant to the orientation of a grain and the alignment of the angular momentum.
  The first subfigure depicts the relationship between the grain body basis ($O_b$) and the angular momentum-based basis ($O_L$). This relationship defines the internal angular momentum alignment, characterised by the Euler angles  $(\phi_b, \theta_b, \psi_b)$.  
  The second subfigure shows the relationship between the $O_L$ and the observer's basis ($O_{\rm obs}$), which defines the external alignment of the angular momentum with respect to the observer, described by the angles  $(\theta_L, \phi_L)$.
  }
  \label{fig: euler angles}
\end{figure}
We use Euler angles to represent the transformations between these three sets of basis vectors, under the convention that:
\begin{equation}
    R_{zxz}(\phi, \theta, \psi)\boldsymbol{v}
    =
    R_{z}(\psi)R_{x}(\theta)R_{z}(\phi)\boldsymbol{v}
\end{equation}
where $\boldsymbol{v}$ stands for an arbitrary spatial vector.
Since the $xy$-axes of $O_L$ can be arbitrary, we take the simplest transformation from $O_{\rm obs}$ to $O_L$: $R_{zxz}(\phi_L, \theta_L, 0)$.
The transformation that rotates $O_L$ to $O_b$ is denoted by $R_{zxz}(\phi_b, \theta_b, \psi_b)$.
These two transformations specify the rotational configuration of a dust grain in the observer's frame. For obvious reasons, we call $(\phi_b, \theta_b, \psi_b)$ the \textit{internal alignment} of the angular momentum, while $(\phi_L, \theta_L)$ is called the \textit{external alignment} of the angular momentum.
The three different coordinate bases and the Euler angles are illustrated in Figure~\ref{fig: euler angles}.

To formalise the torque-free dynamics of the rotating grain, we still need to bridge the angular momentum, the moments of inertia and the angular velocity.
The angular velocity vector $\boldsymbol{\omega}$ represents the rate ($\omega=|\boldsymbol{\omega}|$) at which the object rotates around the direction $\hat{\omega}=\boldsymbol{\omega}/\omega$. Using the convention of Euler angles (from $O_L$ to $O_b$), we can decompose $\omega$ into three successive small angle rotations
\begin{equation}
    \boldsymbol{\omega}
    =
    \Dot{\phi}_b\hat{z}_{L}
    +
    \Dot{\theta}_b \left[R_z(\phi_b)\hat{x}_{L}\right]
    +
    \Dot{\psi}_b
    \left[R_x(\theta_b)R_z(\phi_b)\hat{z}_{L}\right],
\end{equation}
where $\hat{x}_{L}$, $\hat{y}_{L}$ and $\hat{z}_{L}$ are basis vectors of $O_L$.
Applying the transformation $R_{zxz}(\phi_b, \theta_b, \psi_b)$ we can get components of $\boldsymbol{\omega}$ and $\boldsymbol{L}$ on $O_b$, where, taking advantage of the diagonal tensor of inertia, the angular momentum can be represented as
\begin{equation}
    \begin{pmatrix}
        L_1 \\
        L_2\\
        L_3
    \end{pmatrix}_{O_b}
    =
    \begin{pmatrix}
        I_1 & 0 & 0\\
        0 & I_2 & 0\\
        0 & 0 & I_3
    \end{pmatrix}
    \begin{pmatrix}
        \omega_1 \\
        \omega_2\\
        \omega_3
    \end{pmatrix}_{O_b}.
\end{equation}
Here the subscript $O_b$ indicates the basis used here.
After a detailed algebraic manipulation with the rotation matrices, we obtain the following three dynamic equations \cite[see Eq.~(85) in][]{rot_mechanics}
\begin{subequations}
\begin{align}
        \Dot{\theta}_b &= 
        \left(\frac{1}{I_1}
        -
        \frac{1}{I_2}\right)
        L\sin{\theta_b}\sin{\psi_b}\cos{\psi_b},
        \\
        \Dot{\phi}_b &= 
        \left(\frac{\sin^2{\psi_b}}{I_1}
        +
        \frac{\cos^2{\psi_b}}{I_2}\right)L,
        \\
        \Dot{\psi}_b&= \left(\frac{1}{I_3}-\frac{\sin^2{\psi_b}}{I_1}
        -
        \frac{\cos^2{\psi_b}}{I_2}\right)L\cos{\theta_b}.
\end{align}
\end{subequations}
Here $\Dot{\phi}_b$ can be thought of as the \textit{precession} frequency, $\Dot{\theta}_b$ is the \textit{wobble} (or \textit{nutation}) frequency and $\Dot{\psi}_b$ is the \textit{spin} frequency. 
In other words, a torque-free grain has not only the spin state, but also the nutation state, whereas we conventionally refer to rotating dust grains as spinning dust grains.

It also proves to be useful if we reparameterize the principal moments of inertia as below
\begin{align*}
    \frac{1}{I_1}
    &=\frac{1+\alpha}{I_{\rm ref}}, &
    \frac{1}{I_2}
    &=\frac{1-\alpha}{I_{\rm ref}}, &
    \frac{1}{I_3}
    &=\frac{1+\beta}{I_{\rm ref}},
\end{align*}
where we have required 
\begin{equation}
\frac{1}{I_1} - \frac{1}{I_2} = \min\left\{\frac{1}{I_j} - \frac{1}{I_k}\,\Big| \, \text{for all } j,k \text{ satisfying } I_j < I_k\right\}.
\label{eq: tuning grain parameter definition}
\end{equation}
Then, as a result, 
(1) the degeneracy of the different ways of assigning the moments of inertia to the labels $I_1$, $I_2$ and $I_3$ is broken; 
(2) the role of nutation is reduced to a minimum for grain rotation but increased to a maximum for energy distribution.
Furthermore, we call the $xy$-plane (defined with the $O_b$ basis) the ``plane'' and the $z$-axis the ``axis''. Then $\alpha$ characterises the ellipticity of the in-plane structure, and $\beta$ characterises the ratio between the in-plane and axial structure (or say `oblateness'): for $\beta>0$ the grain is more rod-like, for $-1/2<\beta<0$ the grain is more disc-like. For spherical grains $\alpha=\beta=0$, which is the case of \cite{DL98b}. 
And $\alpha=0, \beta= -\frac{1}{2}$ for the oblate grain studied in \cite{SAH11}.
With these definitions the dynamic equations can now be rewritten as
\begin{align}
        \Dot{\theta}_b &= 
        \frac{L}{I_{\rm ref}}
        \alpha
        \sin{\theta_b}\sin{2\psi_b},
        &\Dot{\phi}_b &= \frac{L}{I_{\rm ref}}
        \left(1-\alpha\cos{2\psi_b}
        \right),
        &\Dot{\psi}_b &=
        \frac{L}{I_{\rm ref}}
        \cos{\theta_b}
        \left(\beta+\alpha\cos{2\psi_b}
        \right).
        \label{eq: angular frequecies definition}
\end{align}
By solving the above ODE system, we can obtain the time evolution of the Euler angles and thus the accelerated motion of the electric dipole moment. Then, using the Fourier transform of the electric dipole moment, we can obtain the spectrum of the radiated electric field. We also immediately note that below we shall assume $\alpha=0$ which implies that $\theta_b$ is conserved. This simplifies the modeling considerably, although generalizations can be added.

\begin{figure}[ht]
    \centering

    \begin{minipage}[t]{\textwidth}
        \begin{minipage}[t]{\linewidth}
            \centering
            \begin{minipage}[t]{0.49\linewidth}
                \includegraphics[width=\linewidth]{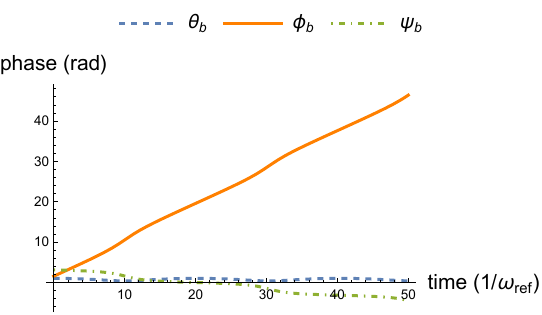}
                \subcaption{Euler angles of the rotating dust grain.}
            \end{minipage}
            \hfill
            \begin{minipage}[t]{0.49\linewidth}
                \includegraphics[width=\linewidth]{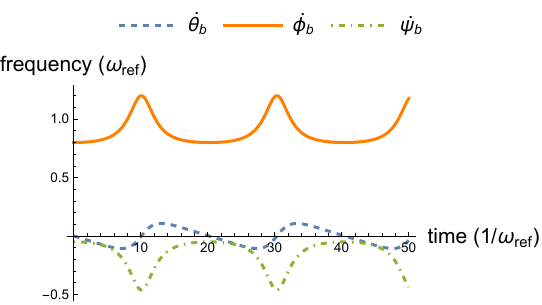}
                \subcaption{Angular frequencies.}
            \end{minipage}
        \end{minipage}

        \vspace{0.3cm} 
        
        \begin{minipage}[t]{\linewidth}
            \centering
            \includegraphics[width=\linewidth]{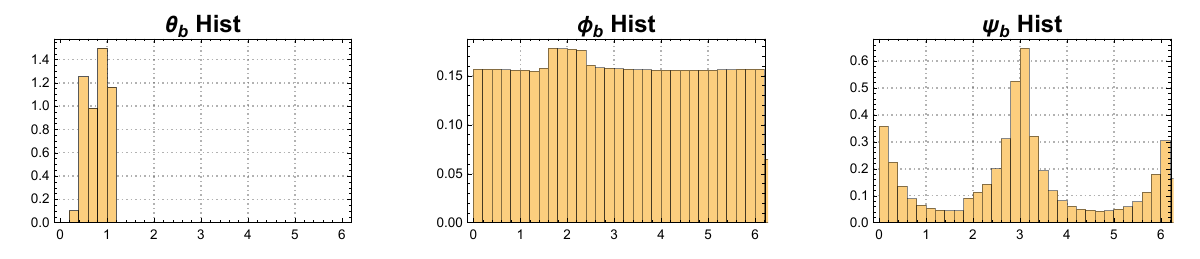}
            \subcaption{Histogram of even-time sampling of grain alignments}
        \end{minipage}

        \vspace{0.3cm} 

        \begin{minipage}[t]{\linewidth}
            \centering
            \begin{minipage}[t]{0.48\linewidth}
                \includegraphics[width=\linewidth]{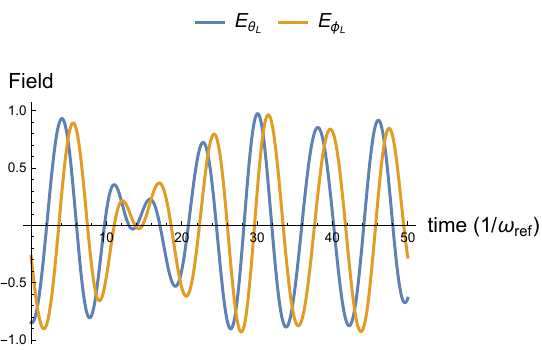}
                \subcaption{$E$ field components (unnormalised).}
            \end{minipage}
            \hfill
            \begin{minipage}[t]{0.48\linewidth}
                \includegraphics[width=\linewidth]{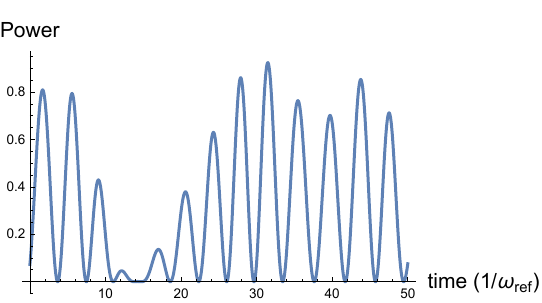}
                \subcaption{Power of the radiation (unnormalised).}
            \end{minipage}
        \end{minipage}

    \end{minipage}
    
    \caption{ Demonstration of torque-free grain dynamics and associated emission mechanisms. 
    (a) Temporal evolution of grain alignment angles (in the grain-body frame).
    (b) Temporal evolution of angular frequencies of the torque-free dust grain. 
    (c) Statistical distribution of Euler angles from uniform time sampling, representing an ensemble of grain rotation configuration when no other processes are considered.
    (d) Radiated electric field components (at an example direction).
    (e) Power intensity at the example direction.
    Grain parameters: $\alpha=0.2$, $\beta=-0.3$. Initial conditions: $\theta_b^0=\pi/3$, $\phi_b^0=\pi/2$, $\psi_b^0=\pi$. 
    The radiation field figures [(d) and (e)] are obtained using a dipole moment with $\mu_1=\mu_2=\mu_3$ and angular moment orientation $\theta_L=\phi_L=\pi/4$.
    The time and frequencies are normalised with the reference frequency:
    $\omega_{\rm ref} \equiv L/I_{\rm ref}$.}
    \label{fig: radiation example}
\end{figure}

{
Having established the dynamics for an arbitrary dust grain shape, we can now formalise the directional radiative electric field and hence the intensity field of the spinning dust grain. A detailed derivation and the higher order correction can be found in the Appendix~\ref{append sec: electrix dipole emission}.
Figure~\ref{fig: radiation example} provides an illustrative example on torque-free grain dynamics and associated emission mechanisms.

We assume that $\alpha$ is approximately zero (with axisymmetry as a special but non-essential case) as in {\tt spdust}, but here we keep $\beta$ as a variable.
Then the radiative electric field is approximately four normal modes, which greatly simplifies the discussion as we can read the power spectrum directly from the normal mode coefficients.
The four modes of the Stokes-$I$ intensity field are
\begin{align}
    \omega^{(1)}&=\Omega, &
    P^{(1)}_\omega &= \frac{1 }{2 } \mu_\parallel^2 \omega^4 \left[3 + \cos(2 \theta_L)\right] \sin^2\theta_b , \\
    \omega^{(2)}&=\Omega \left\vert1+\beta\cos{\theta_b}\right\vert, &
    P^{(2)}_\omega &= 
        \frac{1}{2 } \mu_{\perp}^2\omega^4 \left[3 + \cos(2 \theta_L)\right] \cos^4\left(\frac{\theta_b}{2}\right), \\
    \omega^{(3)}&=\Omega \left\vert1-\beta\cos{\theta_b}\right\vert, &
    P^{(3)}_\omega &= 
        \frac{1}{2 } \mu_{\perp}^2 \omega^4 \left[3 + \cos(2 \theta_L)\right] \sin^4\left(\frac{\theta_b}{2}\right),\\
    \omega^{(4)}&=\Omega\left\vert\beta\cos{\theta_b} \right\vert, &
    P^{(4)}_\omega &= 
        \mu_{\perp}^2 \omega^4 \sin^2 \theta_L \sin^2 \theta_b ,
\end{align}
where $\Omega\equiv {L}/{I_{\rm ref}}$ is the rescaled angular momentum, $\mu_\perp$ is the in-plane ($xy$-plane of $O_b$) component of the electric dipole moment of the dust grain, $\boldsymbol{\mu}$, and $\mu_\parallel$ is the out-of-plane part of $\boldsymbol{\mu}$.
}

\subsection{Toward the overall spectral energy density}
For convenience, the parameters are divided into two types: (1) grain characteristic parameters, $\{I_{\rm ref},\beta,\boldsymbol{\mu}\}$; and (2) rotation configuration parameters: $\{L,\theta_L,\theta_b\}$. 
In addition to these parameters that directly determine the power intensity, there are parameters that describe the environmental conditions. These parameters determine the distribution functions of the grain characteristic parameters and the rotation configuration parameters. 
We refer to these as ``environmental parameters''.

Ideally, if we know the joint distribution of all the parameters, the total spectral energy density (SED) can be obtained as the averaged power intensity over the distribution:\footnote{
Note: (1) We use $f$ to generally represent distribution (or probability) function of variables. When considering the distribution function of a specific subset of variables, omitted variables indicate they have been marginalized out. For example, when discussing the rotation configuration parameters $(L, \theta_L, \theta_b)$ in Section~\ref{sec: config distribution}, we have $f(L) = \int f(L, \theta_L, \theta_b) \, d\theta_L \, d\theta_b$. (2) In the integral form we assume a uniform metric over the differential variables, which means that the distribution function absorbs all metrics and normalisation factors, ensuring that all distribution functions are normalised. Thus, do not worry about the missing $\cos{\theta}/2$ when you see $\diff{\theta}$.}
\begin{multline}
    I_\nu 
    =
    \sum_{m=1}^4 
    \int \diff{L} \diff{\theta_L}
    \diff{I_{\rm ref}} \diff{\beta} \diff{\theta_b}
    \Diff{3}{\boldsymbol{\mu}} 
    \;
    \delta_D\left(\omega - \omega^{(m)}(L, I_{\rm ref}, \beta, \theta_b)\right) \\
    \times 
    P_\omega^{(m)}(\theta_L, \theta_b, \boldsymbol{\mu}) \,
    f(L, \theta_L, I_{\rm ref}, \beta, \theta_b, \boldsymbol{\mu}
    \,\vert\,\textbf{ENV}),
    \label{eq: SED intro form}
\end{multline}
where $m$ denotes the mode, $\omega = 2\pi\nu$, $\delta_D$ is the Dirac delta function, and ``$\textbf{ENV}$'' represents all enviromental parameters. 
These parameters do not directly influence the emission of a grain but do affect the distribution function of other parameters.
Despite conceptual simplicity of Eq.~(\ref{eq: SED intro form}), it is difficult to obtain the full joint distribution because it depends not only on physical but also on environmental processes.
The distribution of rotation configuration parameters can be analysed rationally and deterministically, conditional on environmental parameters and grain characteristics. 
However, the distribution of grain and environmental parameters itself is not known with reasonable certainty and could involve empirical and educated guesses. 

Recognising these difficulties, previous works have considered some distributions that are not physically driven but chosen for mathematical simplicity, such as the log-normal grain size distribution and the Gaussian dipole moment distribution, and limited the discussion to certain grain shapes and environmental conditions.
These treatments greatly facilitate discussion and make forward modelling of the spinning dust emission tractable. 
{\tt SpyDust} must also be explicit about these distribution functions, and its default setting inherits these treatments from {\tt spdust}, although we have made them fully modular, allowing the users to apply their own models and environments as needed, e.g. using a customised distribution function of grain shape and size, $f(I_{\rm ref},\beta)$.

In addition to the use of explicit distribution functions for `forward simulation', there is an opposite logic, akin to `backward fitting'. In this way, one admits ones ignorance on grain and environmental conditions, like the grain size and shape distribution. Perturbative statistical analysis, e.g. moment expansion method \citep{Jens17}, is then used to model the SED with fundamental spectra and moment coefficients. The model is then fit to the data to solve for the coefficients and obtain the SED of the spinning dust.
We present the work done in this logic in a separate paper, but highlight some of the main steps towards realizing this here.

\section{Rotation Configuration Distribution}
\label{sec: config distribution}
The radiation of each grain is determined by its rotational configuration parameters, $\{L,\theta_L,\theta_b\}$, where
$\theta_L$ denotes the external alignment of $\boldsymbol{L}$ in the observer's frame and $\theta_b$ denotes the internal alignment of $\boldsymbol{L}$ in the grain's body frame.
For convenience, we sometimes use the corresponding quantum notations as substitutes: $\{L, \theta_b, \theta_L\} \leftrightarrow \{\ell, m, M\} $.
In this section, we discuss the general form of the distribution function of these rotation configuration parameters.

\subsection{External alignment}
Generally, the configuration space distribution function can be written as
\begin{equation}
    f(L, \theta_b, \theta_L) = f(L, \theta_b | \theta_L)f(\theta_L).
\end{equation}
In this work, we assume an isotropic medium and environment. This assumption leads to an isotropic external angular momentum alignment
\begin{equation}
    f(\theta_L) = \frac{\sin{\theta_L}}{2}.
\end{equation}
It also implies no $\theta_L$ dependence in the rotation configurations of the dust grains, which means that
\begin{equation}
    f(L, \theta_b | \theta_L)
    =
    f(L, \theta_b ).
\end{equation}
Thus, we can rewrite the distribution function of the configuration parameters as
\begin{equation}
    f(L, \theta_b, \theta_L) \approx f(L, \theta_b ) \,
    \frac{\sin{\theta_L}}{2}.
\end{equation}
In the following section we will develop a further breakdown of $f(L, \theta_b)$.
Note that the assumption of isotropic external angular momentum alignment breaks down when we consider the systematic torques imposed by an anisotropic medium or environment, such as a large-scale magnetic field. In this case, one must derive the $\theta_L$ dependent distribution functions.

We note that when considering the thermal equilibrium distribution of angular momentum with a reservoir, we need to take into account that $M$-resolved microstates are fully accessible to energy redistribution.\footnote{We would like to thank Yacine Ali-Ha\"{i}moud for clarifying this aspect.} 
For an isotropic external alignment system, this leads to an apparent `$2\ell+1$'~degeneracy in the distribution of $(\ell, m)$ states.
Again, $\ell, m, M$ are quantum correspondences of $L, \theta_b, \theta_L$.

\subsection{Internal alignment}
The dissipation processes internal to the grain, such as the Barnett effect \cite{barnett1935gyromagnetic}, couple the rotational and vibrational degrees of freedom of the grain. This is known as internal thermal fluctuation and has been well studied in \cite{hoang2010}.
These effects allow rotational energy to be converted into heat and vice versa. If we consider only the internal dissipation process, which preserves the magnitude and the external alignment of the angular momentum, the rotational energy states follow the Boltzmann distribution\footnote{Here, ``conditional on $\ell$'' means that the internal process is approximated to redistribute energy only among $m$ substates with the same $\ell$.}
\begin{equation}
f(m|\ell) \propto \exp{\left(-\frac{E_{\rm rot}(\ell, m)}{kT_{\rm vib}}\right)}, 
\label{eq: boltzmann dist 1}
\end{equation}
where 
$T_{\rm vib}$ is the temperature characterising the internal vibrational energy of the dust grain;
$\ell$ and $m$, which specify the rotational state and energy, are the quantum numbers of the angular momentum magnitude and its projection on the grain $\hat{z}$~axis.
They are quantum equivalents to $L$ and $\theta_b$.
The rotational energy in terms of $\ell$ and $m$ is given by
\begin{equation}
    E_{\rm rot}(\ell, m)
    =
    \frac{
    \left[\ell(\ell+1)
    +
    \beta m^2\right]\hbar^2}{2I_{\rm ref}} .
\end{equation}
Equation~(\ref{eq: boltzmann dist 1}) indicates that the most probable $m$ substate for a grain corresponds to its lowest-energy rotational state. Considering our definition of the grain axes [see Eq.~(\ref{eq: tuning grain parameter definition})], this implies that for $\beta < 0$, the most likely state is $m = \ell$, which corresponds to $\theta_b = 0$, while for $\beta > 0$, the most probable state is $m = 0$, corresponding to $\theta_b = \pi/2$.

In the limit of $E_{\rm rot} \ll kT_{\rm vib}$, which is valid if the rate of absorption of UV photons is larger than the rate of change of the angular momentum, we have
$
    f(m | \ell) = {1}/{(2\ell + 1)},
$
or equivalently in classical notations,
\begin{equation}
    f(\theta_b | L) = \frac{\sin{\theta_b}}{2}.
\end{equation}
This means that the internal vibrational-rotational energy transfer completely randomises the internal alignment.
For small dust grains, this scheme has been shown to be very efficient in the diffuse ISM phases (e.g., CNM, WNM, WIM).
Therefore, we take this isotropic internal alignment assumption as the default setting of {\tt SpyDust}. 
As a result of the isotropic internal alignment, we then have 
\begin{equation}
    f(L, \theta_b)
    =
    f(\theta_b | L) f(L)
    =
    f(\theta_b ) f(L)
    =
    \frac{\sin{\theta_b}}{2} f(L),
\end{equation}
from which we also deduce $f(L|\theta_b) = {f(L,\theta_b)}/{f(\theta_b)} = f(L)$.
The full distribution function of the rotation configuration parameters can then be rewritten as follows:
\begin{equation}
    f(L, \theta_b, \theta_L) 
    = f(L) f(\theta_b )  f(\theta_L )
    =  f(L) \frac{\sin{\theta_b}}{2}
    \frac{\sin{\theta_L}}{2}.
\end{equation}
This factorisation greatly simplifies the subsequent averaging process. We reiterate that if the assumptions of random internal alignment or external alignment are not met, the factorisable form presented above is no longer valid, and corrections are expected.

\subsection{Angular momentum distribution}
In this section, we discuss the distribution function of the angular momentum magnitude, $f(L)$. First, we consider the dissipation and fluctuation processes of the angular momentum vector, $\boldsymbol{L}$. Ultimately, the distribution $f(\boldsymbol{L})$ is determined by the sum of all dissipation vectors and the sum of all fluctuation tensors.
It is important to emphasize that we only need to consider the dissipation and fluctuation of the $\boldsymbol{L}$ states, rather than the combined $(\boldsymbol{L}, \theta_b)$ states, due to the isotropic internal alignment previously discussed (see Section~\ref{sec: fokker planck} for further details).

\subsubsection{Angular momentum dissipation and fluctuation}
\label{sec: fokker planck}
The evolution of the rotation configuration distribution $f(\boldsymbol{L}, \theta_b)$ is given by the Boltzmann equation of the following Ansatz\footnote{
    Although we started with this Ansatz, it's important not to take it for granted. Starting from the primitive point, specifically the ``Vlasov equation'' (the terminology ``Vlasov equation'' is used loosely here, as we introduce an non-canonical variable $\theta_b$), terms such as the derivatives $\dot{L}$ and $\dot{\theta}_b$ appear. The $\dot{\theta}_b$ term vanishes due to our assumption that $\alpha = 0$, while $\dot{L}$ vanishes due to this argument: The intrinsic phase space is defined in terms of angular momentum and conjugate angles, but we effectively marginalise over the angular variables. Since these are canonical classical variables, the divergence of the ``phase space velocity'' field is zero, resulting in the disappearance of $\dot{L}$ terms.}
    \begin{multline}
        \frac{\partial f(\boldsymbol{L}, \theta_b)}{\partial t}
        =
        \int 
        \;
        \Big[
        f(\boldsymbol{L}+\Delta \boldsymbol{L}, \theta_b+\Delta\theta_b)
        \Gamma(\boldsymbol{L}+\Delta \boldsymbol{L}\rightarrow \boldsymbol{L}, \theta_b+\Delta\theta_b\rightarrow\theta_b) \\
        -
        f(\boldsymbol{L}, \theta_b)
        \Gamma(\boldsymbol{L}\rightarrow \boldsymbol{L}+\Delta \boldsymbol{L},
        \theta_b\rightarrow\theta_b+\Delta\theta_b
        )
        \Big]
        \Diff{3}{\!\Delta\! \boldsymbol{L}}\diff{\Delta\theta_b},
    \end{multline} 
    where $\Gamma$ is the transition rate of the grain rotational states which is determined by the environmental conditions.
    The equation basically says that the rate of change of the number of particles in the $(\boldsymbol{L},\theta_b)$ state is equal to all ``incoming'' particles minus all ``outgoing'' particles.

    Due to the assumption of sufficiently efficient internal thermal fluctuations - meaning the timescale of $\Delta \theta_b$ is much shorter than that of $\Delta L$ - and the condition $T_{\rm vib} \gg T_{\rm rot}$, we adopt an isotropic $\theta_b$ distribution, expressed as $f(\boldsymbol{L}, \theta_b) = (\sin \theta_b / 2) f(\boldsymbol{L})$. This allows us to bypass an explicit treatment of the internal processes governing the distribution of $\theta_b$.\footnote{In other words, one will also have to consider the dissipation and fluctuation of $\theta_b$ when the assumption of isotropic internal alignment breaks down.}
    As a result, the transition equation is simplified to the following form
    \begin{equation}
    \begin{split}
        \frac{\partial f(\boldsymbol{L}) }{\partial t} 
        &=
        \int 
        \Diff{3}{\!\Delta\! \boldsymbol{L}}
        \left[
        f(\boldsymbol{L}+\Delta \boldsymbol{L})
        \Big\langle\Gamma(\boldsymbol{L}+\Delta \boldsymbol{L}\rightarrow \boldsymbol{L}; \theta_b)\Big\rangle_{\theta_b}
        -
        f(\boldsymbol{L})
        \Big\langle
        \Gamma(\boldsymbol{L}\rightarrow \boldsymbol{L}+\Delta \boldsymbol{L}; \theta_b
        )\Big\rangle_{\theta_b}
        \right],
    \end{split}
    \label{eq: transition equation averaged over theta}
    \end{equation}
    where $\Gamma$ has been redefined as the $\theta_b$-dependent transition rate of the $\boldsymbol{L}$ states, rather than the transition rate of the combined $(\boldsymbol{L}, \theta_b)$ states, and $\langle{\cdots}\rangle_{\theta_b}$ represents the ensemble averaging over $\theta_b$.
    For convenience in later discussions, we define $\Bar{\Gamma}$ as the value of $\Gamma$ averaged over $\theta_b$.

    For a distribution function in equilibrium, we have \citep{ali13review}
    \begin{equation}
        \int f(\boldsymbol{L})\Bar{\Gamma}(\boldsymbol{L}\rightarrow \boldsymbol{L}') \Diff{3}{\boldsymbol{L}'}
        =
        \int f(\boldsymbol{L}')\Bar{\Gamma}(\boldsymbol{L}'\rightarrow \boldsymbol{L}) \Diff{3}{\boldsymbol{L}'}.
    \end{equation}
    If we further assume that the change in angular momentum, $\Delta\boldsymbol{L}=\boldsymbol{L}' - \boldsymbol{L}$, is small, we can safely perform a Taylor expansion of $f$ and $\Gamma$ around $\boldsymbol{L}'$. This leads to the Fokker-Planck equation
    \begin{equation}
        \frac{\partial}{\partial L^i}
        \left[
        \Bar{D}^i(\boldsymbol{L}) f(\boldsymbol{L})
        \right]
        =
        \frac{1}{2}
            \frac{\partial^2}{\partial L^i\partial L^j}
            \left[
            \Bar{F}^{ij}(\boldsymbol{L})f(\boldsymbol{L})
            \right],
    \end{equation}
    where $\Bar{D}^i$ is the dissipation vector averaged over $\theta_b$, defined as
    \begin{align}
        \Bar{D}^i(\boldsymbol{L})
        &
        =\langle D^i(\boldsymbol{L}, \theta_b) \rangle_{\theta_b},
        &
        D^i(\boldsymbol{L}, \theta_b)
        &\equiv\frac{\diff {\langle{\Delta L^i}\rangle}}{\diff{t}}\Big\vert_{\boldsymbol{L}, \theta_b}
        =
        \int 
            \Delta L^i
            \Gamma(\boldsymbol{L}\rightarrow\boldsymbol{L}
            +\Delta\boldsymbol{L}; \theta_b) 
            \Diff{3}{\!\Delta\! \boldsymbol{L}},
    \end{align}
    and $\Bar{F}^{ij}$ is the fluctuation tensor averaged over $\theta_b$:
    \begin{align}
        \Bar{F}^{ij}(\boldsymbol{L})
        &=\langle F^{ij}(\boldsymbol{L}, \theta_b) \rangle_{\theta_b},
        &
        F^{ij}(\boldsymbol{L}, \theta_b)
        &\equiv
        \frac{\diff {\langle{\Delta L^i \Delta L^j}\rangle}}{\diff{t}}\Big\vert_{\boldsymbol{L}, \theta_b}
        =
        \int 
            \Delta L^i \Delta L^j
            \Gamma(\boldsymbol{L}\rightarrow\boldsymbol{L} + \Delta\boldsymbol{L}; \theta_b) 
            \Diff{3}{\!\Delta\! \boldsymbol{L}}.
    \end{align}
    Here, the notation ``$\cdots\big\vert_{\boldsymbol{L}, \theta_b}$'' indicates that the rates are evaluated for particles prepared in the state of $(\boldsymbol{L}, \theta_b)$. 

\subsubsection{Fokker-Planck equation for angular momentum magnitude}
The change in the angular momentum, $\boldsymbol{L}\rightarrow\boldsymbol{L}+\Delta\boldsymbol{L}$, results in a change in the magnitude,
        \begin{equation}
            \Delta L \equiv |\boldsymbol{L}+\Delta\boldsymbol{L}| - |\boldsymbol{L}|, 
        \end{equation}
        and generally $\Delta L \neq |\Delta \boldsymbol{L}| $.
        We can decompose the magnitude ``drift'' as a sum of spatial components drifts by expanding $\Delta L$ in terms of $L^i$ and $\Delta L_i$:
        \begin{equation}
            \Delta L
            \simeq
            \frac{L_i\Delta L^i}{L}
            +\frac{1}{2} \frac{\Delta L_i\Delta L^i}{L}
            -\frac{1}{2} \frac{L_iL_j\Delta L^i\Delta L^j}{L^3} .
        \end{equation}
        This expansion can be used to find the dissipation rate for the magnitude of the angular momentum for a particle prepared in the state $\boldsymbol{L}$ 
        \begin{equation}
            D_L(\boldsymbol{L}) 
            \equiv
            \frac{\diff {\langle \Delta L \rangle}}{\diff{t}}\Big\vert_{\boldsymbol{L}}
            \simeq 
            \frac{1}{L} 
            \left[
            L_i\Bar{D}^i
            +
            \frac{1}{2} \Bar{F}^i_{\;\;i}
            -
            \frac{1}{2L^2} L_iL_j\Bar{F}^{ij}
            \right].
            \label{eq: magnitude damping rate}
        \end{equation}
        Similarly, the excitation rate of the angular momentum magnitude is given by
        \begin{equation}
            {F}_L(\boldsymbol{L})
            \equiv\frac{\diff {\langle{(\Delta L)^2}\rangle}}{\diff{t}}\Big\vert_{\boldsymbol{L}}
            \simeq\frac{1}{L^2} L_iL_j\Bar{F}^{ij}.
            \label{eq: magnitude fluctuation rate}
        \end{equation}
        Since the medium and the angular momentum orientation are isotropic, the rates are independent of the angular momentum orientation: $D_L(L) = D_L(\boldsymbol{L}) $ and $F_L(L) = F_L(\boldsymbol{L}) $.
        Taking advantage of the $O_L$ basis where $\boldsymbol{L}=L\hat{z}$, the magnitude dissipation and fluctuation rates are simplified to \citep{ali13review}
    \begin{align}
        D_L&= 
            \Bar{D}^z +
            \frac{1}{2L} \left(\Bar{F}^{xx}+\Bar{F}^{yy}\right),
        &
        F_L&= \Bar{F}^{zz}.
    \end{align}
        To study the steady distribution of the angular momentum magnitude $f(L)$, we can write down the Fokker-Planck equation as follows
\begin{equation}
        \frac{\partial}{\partial L}
        \left[
        D_L(L) f(L)
        \right]
        =
        \frac{1}{2}
        \frac{\partial^2}{\partial L^2}
            \left[
            F_L(L)
            f(L)
            \right],
        \label{eq: magnitude Fokker Planck}
\end{equation}
which can be used to compute $f(L)$.

{
Due to the linearity of Eq.~(\ref{eq: magnitude damping rate}) and (\ref{eq: magnitude fluctuation rate}), we can simply sum the magnitude dissipation and fluctuation rates across processes:
    \begin{align}
        D_L &= \sum_X D_L^{(X)},
        &
        F_L = \sum_X F_L^{(X)},
    \end{align}
where $X$ indexes the angular momentum transport scheme. 
This synthesis of different processes is the core strategy of \cite{AHD09} and \cite{SAH11}. 
Assuming a homogeneous solution form, one can solve for the distribution function $f(L)$ given the dissipation rate $D_L$ and the fluctuation rate $F_L$. See Appendix~\ref{append: use of FK} for the use of the Fokker-Planck equation.
}

\subsection{Radiative damping}
\label{sec: radiative damping}
In Section~\ref{sec: dynamics} we have discussed the torque-free dynamics and radiation of the dust grains. 
``Torque-free'' is an ideal assumption that holds on the time scale of the coherence time of the radiation modes. 
However, due to the back-reaction of the radiation, even a dust grain isolated from the environment will lose its angular momentum over time. This effect is known as {\it radiative damping}.
Next, we derive the angular momentum dissipation caused by this mechanism. Our equation for the electric dipole back-reaction is a correction and generalization of the relevant expression in {\tt spdust}.

First of all, we revisit the assumption of a constant electric dipole moment that we made when discussing the torque-free rotation of the grain. This assumption is not exactly correct because of the vibrations of the chemical bonds within the grain body. 
More precisely, we rewrite the electric dipole moment as 
\begin{equation}
    \boldsymbol{\mu}_{\rm true}
    =
    \boldsymbol{\mu}
    +
    \delta\boldsymbol{\mu},
\end{equation}
where $\boldsymbol{\mu}$ is the component fixed to the rotating grain, called the intrinsic dipole moment; while $\delta\boldsymbol{\mu}$ is the deviation of $\boldsymbol{\mu}$ from the true electric dipole moment, taking into account the internal vibrations.
The radiation power can be roughly understood as the sum of the contributions from the two parts,
$P_{\omega}
    \sim
    \langle{\Ddot{\mu}}^2\rangle
    +
    \langle{\delta\Ddot{\mu}}^2\rangle$.
In terms of the AME spectrum, the $\delta\boldsymbol{\mu}$ contribution can be safely discarded, as it has a bigger effect on the infrared emission, which is much higher in frequency than the frequencies at which the grain rotates. 
We also assumed that $\langle{\Ddot{\mu}\delta\Ddot{\mu}} \rangle=0$.

\begin{figure}
    \centering
    \includegraphics[width=\linewidth]{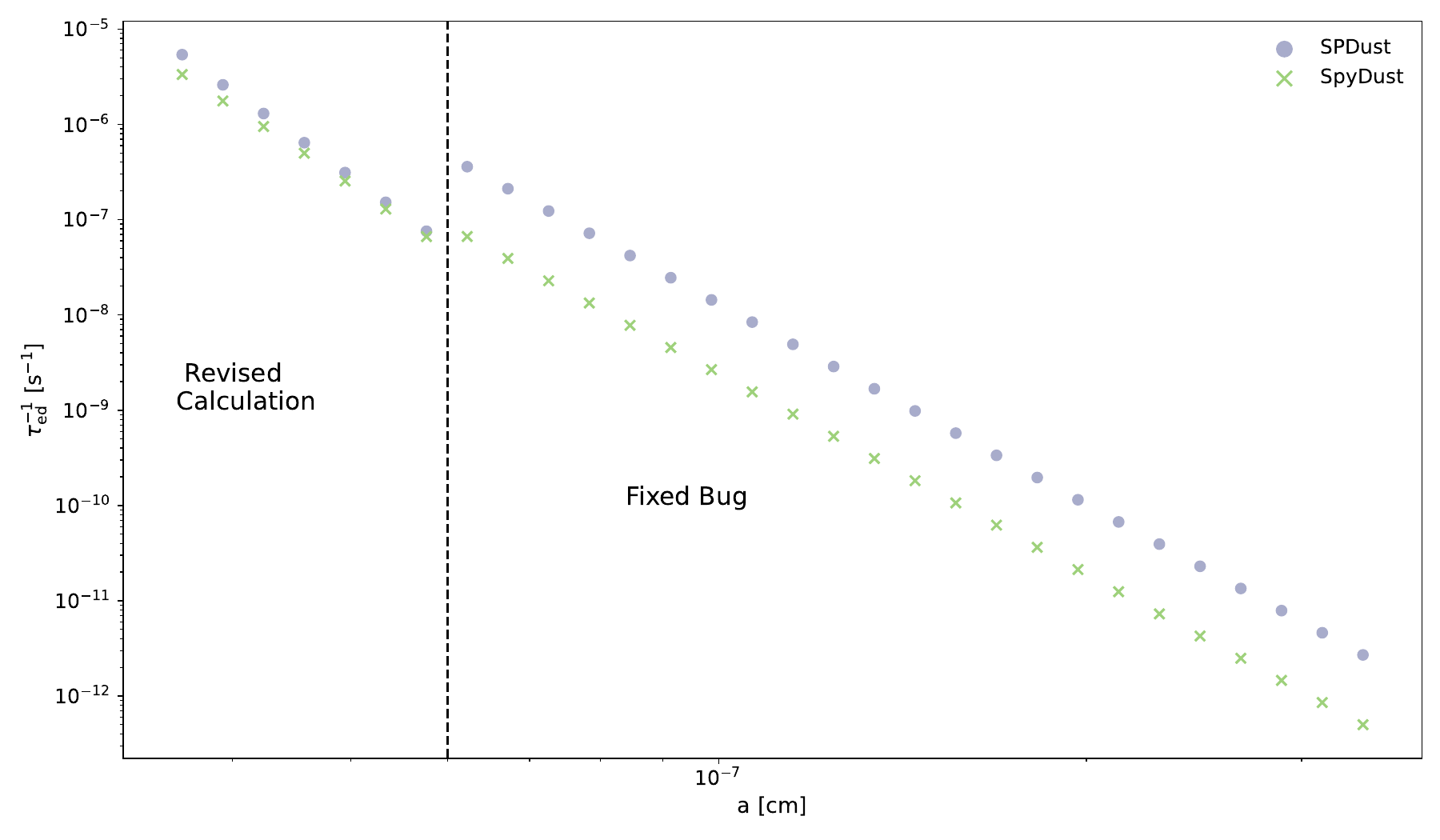}
    \caption{Comparison of the inverse characteristic damping time,  $\tau_{\rm ed}^{-1}$, due to radiation backreaction in {\tt spdust}, shown for different grain sizes and shapes. The vertical dashed line (at $6\times 10^{-8}$~cm) divides disc-like grains (left) from spherical grains (right). 
    This example uses the WNM environment defined in \cite{DL98b}, Table~1. 
    On the left of the dashed line: 
    The plot shows the inverse characteristic time after correcting the $\beta$ values of disc-like grains.
    On the right of the dashed line: 
    The corrected calculation resolves a bug in {\tt spdust} that misapplied the ``tumbling'' setting, causing it to override the shape distinction between spherical and non-spherical grains.}
    \label{fig:tau_ed_inv}
\end{figure}

Although the radiation from $\delta\mu$ does not contribute directly to the bands of interest, it transports the angular momentum, so that it indirectly affects the SED of the spinning dust emission.
The vector rate of the angular momentum drift is given by 
\begin{equation}
    \boldsymbol{D}(\boldsymbol{L}, \theta_b) 
    \equiv
    \left\langle\frac{\diff{\Delta\boldsymbol{L}}}{\diff{t}}\right\rangle \Big\vert_{\boldsymbol{L},\theta_b}
    =
    -\frac{2}{3c^3} \langle{\Dot{\boldsymbol{\mu}}_{\rm true}\times\Ddot{\boldsymbol{\mu}}_{\rm true}}\rangle
    =
    -\frac{2}{3c^3} \left(\langle \Dot{\boldsymbol{\mu}}\times\Ddot{\boldsymbol{\mu}} \rangle+\langle{\delta\Dot{\boldsymbol{\mu}}\times\delta\Ddot{\boldsymbol{\mu}}}\rangle\right).
\end{equation}
The second term, the radiative damping due to infrared photons, is well established \citep{SAH11} and will not be repeated here. 
The first term is calculated by {\tt spdust} for the case of $\beta=-1/2$.
Below we revisit the first term, the dissipation (or drift) rate of the angular momentum caused by the back-reaction of the electric dipole emission:
\begin{equation}
    \boldsymbol{D}_{\rm (ed)}(\boldsymbol{L}, \theta_b) 
    \equiv 
    -\frac{2}{3c^3} \langle{\Dot{\boldsymbol{\mu}}\times\Ddot{\boldsymbol{\mu}}}\rangle.
\end{equation}
This ensemble average can be evaluated over rotation periods of $\psi$ and $\phi$, which gives
        \begin{equation}
        D^i_{\rm (ed)}(\boldsymbol{L}, \theta_b)
            =-\frac{\mu_\perp^2 \left[1 + \left(1 + 3 \beta (2 + \beta)\right) \cos^2\theta_b + \beta^2 (3 + 2 \beta) \cos^4\theta_b\right] + 2 \mu_\parallel^2 \sin^2\theta_b}{3 c^3 I_{\rm ref}^3}
           \, L^2 L^i ,
        \end{equation}
where the superscript $i=x,y,z$ denotes the spatial component of the vector.
The dissipation rate averaged over internal alignments is then given by
        \begin{equation}
        \Bar{D}^i_{\rm (ed)}(\boldsymbol{L})
            = \int D^i_{\rm (ed)}(\boldsymbol{L}, \theta_b) \,\frac{\sin(\theta_b)}{2} \,{\rm d} \theta_b
            =-\frac{L^2 L^i}{3 c^3 I_{\rm ref}^3}\left[\mu_\perp^2 \left( 
            \frac{2}{5}\beta^3 +
            \frac{8}{5}\beta^2 +
            2\beta + \frac{4}{3} 
            \right) + \frac{4}{3} \mu_\parallel^2 \right] ,
        \end{equation}
where the bar represents the average over $\theta_b$.

In line with the convention in {\tt spdust}, we define the characteristic damping time for angular momentum drift due to radiation backreaction as:
\begin{equation}
    \tau_{\rm ed}^{-1}
    \equiv
    -\frac{3I_3 k T}{L^3} 
    \left\langle\frac{\diff{\Delta L}}{\diff{t}}\right\rangle
    =
    \frac{k_B T}{I_3^2 c^3 (1+\beta)^3}
    \left[\mu_\perp^2 \left( 
            \frac{2}{5}\beta^3 +
            \frac{8}{5}\beta^2 +
            2\beta + \frac{4}{3}
            \right) + \frac{4}{3} \mu_\parallel^2 \right].
\end{equation}
For $\beta=-1/2$, it reduces to $\tau_{\rm ed}^{-1}=\frac{3kT}{I_3^2 c^3}\left(\frac{82}{45}\mu_\perp^2 + \frac{32}{9}\mu_\parallel^2 \right)$, which exactly matches the Eq.~(48) of \cite{SAH11}. However, $\beta=-1/2$ is inaccurate for general disc-like grains: for the {\tt spdust} grains with $a < 6\text{ \AA}$ and a thickness about the typical separation between graphene layers (which is approximated by {\tt spdust}), $\beta$ actually ranges from $-0.47$ to $-0.39$.\footnote{This can be evaluated directly using Eq.~(\ref{eq: size parameter of elliptical cylinder grain}).} 
Figure~\ref{fig:tau_ed_inv} shows a comparison of  $\tau_{\rm ed}^{-1}$ before and after the correction to the $\beta$ value of disc-like grains.
While the formula for spherical grains (corresponding to the right side of the vertical dashed line) remains unchanged, we have corrected a subtle bug in {\tt spdust} where the setting of the ``tumbling''\footnote{In {\tt spdust}, ``tumbling'' refers to an isotropic internal orientation.} keyword inadvertently overrode the distinction between spherical and non-spherical dust grains.

\subsection{Plasma Effect on Spinning Dust Grains}
\label{sec: plasma}

In the following we will first calculate the rates of angular momentum fluctuation caused by the plasma drag. Then with $f(L)$ and $F_L$ in hand, we solve for the dissipation rate $D_L$ using Eq.~\eqref{eq: dissipation rate from Fokker-Planck}.
Since this section frequently refers to the parameters $\theta_b$, $\phi_b$ and $\psi_b$, which describe internal alignment, we have omitted the subscript for the sake of brevity. 
Below we present the angular momentum dissipation and fluctuation rates caused by plasma effects, generalised to arbitrary $\beta$, with the detailed derivation and discussion in the Appendix~\ref{append: plasma}.

The $\theta$-average of the fluctuation rate of the magnitude is given by
\begin{equation}
    F_L = \Bar{F}^{zz}
    =
    \frac{\mu_\perp^2}{2}
    \int_{-1}^1(1-x)^2 
    P_E\left(\nu_{\rm ref} (1-\beta x)\right) \diff{x}
     \\
    + \frac{2}{3}\,\mu_\parallel^2 P_E\left(\nu_{\rm ref}\right),
    \label{eq: fluctuation plasma L}
\end{equation}
where we substituted $x=\cos\theta$.
The cross-fluctuation between the angular momentum amplitude and its projection onto the grain axis, $\hat{z}_b$, is expressed as:
\begin{equation}
    F^{zz_b}
    \equiv
    \frac{\diff{\braket{\Delta L \, \Delta L_{z_b}}}}{\diff{t}}
    =
    {\mu_\perp^2}  \left[
    \frac{(1+\cos{\theta})^2}{4}P_E\left(\nu_{\rm ref} (1+\beta\cos{\theta})\right)
    -\frac{(1-\cos{\theta})^2}{4} 
    P_E\left(\nu_{\rm ref} (1-\beta\cos{\theta})\right)
    \right] 
\end{equation}
with $\Delta L_{z_b} = \Delta L\cos{\theta}$.
The fluctuation rates driven by the stochastic ambient plasma electric field are formalised as a $\beta$-generalised extension of the calculations in \cite{SAH11}. These rates are then used to derive the angular momentum drift rate. 


%

To better align with the formalism in {\tt spdust}\footnote{See Appendix~\ref{app:derivation_Fpl} for the conversions between the notations.}, we rescale the plasma dissipation and fluctuation rates following the {\tt spdust} convention, where the fluctuation rate of the angular momentum magnitude is rewritten as a dimensionless coefficient
\begin{equation}
    \mathcal{G}_{\rm pl}(\Tilde{\Omega})
    =\frac{I_3\tau_{\rm H}}{2kT}
    \frac{F_L(\boldsymbol{L})}{I_3^2},
\end{equation}
where $\Tilde{\Omega}\equiv \frac{L}{I_3}$,
and $\tau_{\rm H}$ is the idealized damping timescale defined in \cite{AHD09}.
By substituting Eq.~(\ref{eq: fluctuation plasma L}), the specific implementation of $\mathcal{G}_{\rm pl}$ in {\tt SpyDust} is given by:
\begin{multline}
\label{eq:def_G_pl}
    \mathcal{G}_{\rm pl, {\tt SpyDust}}(\Tilde{\Omega},\beta)
    =
    \left\langle
    \frac{(1-\cos{\theta})^2}{2}\, 
     G_{\rm pl, AHD}\left(\Tilde{\Omega} \frac{1-\beta\cos{\theta}}{1+\beta}\right) 
     \right\rangle_{\theta}
     + 
     \frac{2}{3}
     \frac{\mu_\parallel^2}{\mu_\perp^2}\, G_{\rm pl, AHD}\left(\frac{\Tilde{\Omega}}{1+\beta} \right),     
\end{multline}
where we have used the power spectrum calculation of the plasma electric field as explained in \cite{SAH11}:
\begin{equation}
    P_E(\nu) = \frac{2I_3 k T}{\tau_{\rm H} \mu_\perp^2} G_{\rm pl, AHD} (2\pi \nu),
\end{equation}
with $G_{\rm pl, AHD}$ denoting the plasma-induced fluctuation rate calculated in \cite{AHD09}.
By transforming variables, we can further simplify Eq.~\eqref{eq:def_G_pl} to
\begin{align}
\mathcal{G}_{\rm pl}(\Tilde{\Omega},\beta)
    &=
    - \frac{(1+\beta)^3}{4\beta^3} \!\int_{\Tilde{\Omega}}^{\frac{1-\beta}{1+\beta}\Tilde{\Omega}}
    \left(\frac{1-\beta}{1+\beta}-\frac{\omega}{\Tilde{\Omega}}\right)^2
    G_{\rm pl, AHD}\left(\omega\right)\frac{{\rm d}\omega}{\,\Tilde{\Omega}}
    +
    \frac{2}{3}
     \frac{\mu_\parallel^2}{\mu_\perp^2}\, G_{\rm pl, AHD}\left(\frac{\Tilde{\Omega}}{1+\beta} \right)
     \nonumber \\[1mm]
     &\!\!\!\!\!\!
     \stackrel{\beta = -1/2}{\stackrel{\downarrow}{=}}
    \frac{1}{4} \int_{\Tilde{\Omega}}^{3\Tilde{\Omega}}\,
    \left(3-\frac{\omega}{\Tilde{\Omega}}\right)^2\,
    G_{\rm pl, AHD}\left(\omega\right)\frac{{\rm d}\omega}{\Tilde{\Omega}}
    +
\frac{2}{3}
     \frac{\mu_\parallel^2}{\mu_\perp^2}\, G_{\rm pl, AHD}\left(2\Tilde{\Omega} \right),   
\end{align}
which precisely matches the {\tt spdust} treatment for $\beta = -1/2$ \citep[see Eq.~(91) of][]{SAH11}, but here is given for general $\beta$.
Also following the convention in \cite{SAH11}, we write the detailed balance-derived dissipation rate due to plasma effects as a dimensionless coefficient in the form \citep[cf. Equation (86) and (92) in][]{SAH11}
\begin{equation}
\begin{split}
    \mathcal{F}_{\rm pl} &= \frac{\tau_{\rm H}}{\Tilde{\Omega}} 
    \frac{I_3 \Tilde{\Omega}}{2 k T_{\rm pl}}
    \left[
        \frac{I_3}{I_{\rm ref}} \frac{\langle F^{zz}\rangle_\theta}{I_3^2}
        -
        \frac{I_3-I_{\rm ref}}{I_{\rm ref}}\frac{\langle \cos{\theta} F^{zz_b} \rangle_\theta}{ I_3^2}
    \right] \\
    &= 
    \frac{\mathcal{G}_{\rm pl}(\Tilde{\Omega}) }{1+\beta}
    + \frac{1}{2} \frac{\beta}{1+\beta} \int_{-1}^1
    \Bigg\{
    \frac{(1+x)^2}{4} G_{\rm pl, AHD}\left[\left(\frac{1+\beta x}{1+\beta}\right)\Tilde{\Omega}\right]
    -\frac{(1-x)^2}{4} G_{\rm pl, AHD}\left[\left(\frac{1-\beta x}{1+\beta}\right)\Tilde{\Omega}\right]
    \Bigg\} \,x\diff{x} 
\end{split}
\end{equation}
which, after transforming the variables, can be rewritten as
\begin{equation}
\begin{split}
    \mathcal{F}_{\rm pl}(\Tilde{\Omega}) 
    &= 
    - \frac{(1+\beta)^3}{4\beta^3} \!\int_{\Tilde{\Omega}}^{\frac{1-\beta}{1+\beta}\Tilde{\Omega}} \frac{\omega}{\Tilde{\Omega}}
    \left(\frac{1-\beta}{1+\beta}-\frac{\omega}{\Tilde{\Omega}}\right)^2
    G_{\rm pl, AHD}\left(\omega\right)\frac{{\rm d}\omega}{\,\Tilde{\Omega}} 
    +  \frac{2}{3(1+\beta)}
     \frac{\mu_\parallel^2}{\mu_\perp^2}\, G_{\rm pl, AHD}\left(\frac{\Tilde{\Omega}}{1+\beta} \right)
    \\
    &\!\!\!\!\!\!
     \stackrel{\beta = -1/2}{\stackrel{\downarrow}{=}} 
    \frac{1}{4}\int_{\Tilde{\Omega} }^{3\Tilde{\Omega}}
    \frac{\omega}{\Tilde{\Omega}}
    \left(3-\frac{\omega}{\Tilde{\Omega}}\right)^2
    G_{\rm pl,AHD}(\omega)\frac{\diff{\omega}}{\Tilde{\Omega}}
    +
    \frac{4\mu_{\parallel}^2}{3\mu_{\perp}^2}
    G_{\rm pl,AHD}(2\Tilde{\Omega}).
\end{split}
\end{equation}
This also matches the {\tt spdust} result of $\mathcal{F}_{\rm pl}$ for $\beta=-1/2$.


\newpage
\section{Spectral Energy Density}
\label{Sec: SED}
In this paper we use $I_{\rm ref}$, $\alpha$ and $\beta$ to represent the moments of inertia of the grain.
These are parameters that directly affect the electrical dipole radiation. In addition to these parameters, we also need to clarify the grain geometry, which affects the effective surface area of the grain, the dipole moment distribution and some other properties, and therefore is one of the rotational distribution parameters.

Unlike the two typical grain shapes used in {\tt spdust} - where larger grains are assumed to be spherical and smaller grains are modelled as discs - we consider two more general shapes: an ellipsoid for larger grains and an elliptical cylinder for smaller grains. This approach makes the grain shapes in {\tt spdust} a special case within the framework of {\tt SpyDust}.
For a direct link to \cite{AHD09} and \cite{SAH11}, we define the volume-equivalent radius, $a$, which is an alternate grain size parameter with a similar role to $I_{\rm ref}$. We formalise $a$ in terms of $I_{\rm ref}$ and $\beta$ in the section~\ref{sec: volume equivalent radius a}.

\subsection{Hierarchical ensemble averaging}
\label{sec: Hierarchical ensemble averaging}
From the discussion in section~\ref{sec: config distribution} we can write the distribution function of the rotation configuration parameters as
\begin{equation}
     f(\Omega, \theta_L, a, \beta, \theta_b, \boldsymbol{\mu})
     =
     f(\Omega, a, \beta, \boldsymbol{\mu})
     f(\theta_L)f(\theta_b),
\end{equation}
where $\Omega\equiv L/I_{\rm ref}$, whose distribution is given by $f(\Omega, I_{\rm ref}, \beta, \boldsymbol{\mu})=f(L, I_{\rm ref}, \beta, \boldsymbol{\mu})(\diff{L}/\diff{\Omega})$, and we have used the grain size parameter $a$ instead of $I_{\rm ref}$.
We build on previous approaches to discussing angular momentum distribution, which can be summarized in the following form:
\begin{equation}
\begin{split}
    f(\Omega, a, \beta, \boldsymbol{\mu})
    =
    f(\Omega | a, \beta, \boldsymbol{\mu})
    f(\boldsymbol{\mu} | a, \beta)
    f(\beta | a) f(a)
\end{split}
\end{equation}
In other words, we assume a prior grain size distribution. For grains of different sizes, we consider different shapes: specifically, for $a > 6 \ \text{\AA}$ , an ellipsoid shape, and for $a \leq 6 \ \text{\AA}$, an elliptical cylinder shape. Additionally, {\tt SpyDust} allows for the consideration of a distribution over the shape parameter $\beta$. Then, for a given grain size and shape, we specify the distribution of electric dipole moments and calculate the resulting angular momentum distribution.

By leveraging the factorizable form of the distribution function and the specific dependencies within each component, we can express the overall SED as:
\begin{multline}
    I_\nu 
    = 
    \int \diff{\Omega} \diff{\beta} 
    \Diff{3}{\boldsymbol{\mu}} \,
    \left(
    \int \diff{a} \,
    f(\Omega, a, \beta, \boldsymbol{\mu}) 
    \right)
    \Bigg[ \sum_{m=1}^4
    \int\diff{\theta_b}\, \delta_D\left(\omega - \omega^{(m)}(\Omega, \beta, \theta_b)\right)\\
    \times 
    f(\theta_b) \left(\int\diff{\theta_L}\,f(\theta_L)P_\omega^{(m)}(\theta_b, \theta_L, \boldsymbol{\mu}) \right)
    \Bigg]
\end{multline}
This formulation separates the integration over each parameter while preserving the dependencies of each function, allowing a systematic evaluation of the spectrum. The Dirac $\delta$ function, $\delta_D(\omega - \omega^{(m)}(\Omega, \beta, \theta_b))$, constrains the frequency $\omega$ for each mode $m$, and the nested integrations further refine the contributions based on their respective distributions.

The Dirac $\delta$ functions effectively eliminate an integral, depending on which variable ($\Omega$, $\beta$ or $\theta_b$) you choose to evaluate the Dirac $\delta$ functions against.
In abstract terms, if we choose to eliminate the $\theta_b$ integral, we are left with integrals of the following form
\begin{equation}
\begin{split}
    I_\nu 
    =
    \int \diff{\Omega}
    \diff{\beta} 
    \Diff{3}{\boldsymbol{\mu}} \,
    \frac{
    P_\omega\left(\theta_b(\Omega, \beta), \boldsymbol{\mu}\right) }{2|\partial\omega/\partial\cos{\theta_b}|} \,
    f(\Omega, \beta, \boldsymbol{\mu}),
    \label{eq: dirac delta on theta}
\end{split}
\end{equation}
where $P_\omega$ abstractly denotes the power related term.
On the other hand, we can also evaluate the Dirac $\delta$ function with $\Omega$, which yields
\begin{equation}
    I_\nu 
    =
    \int 
    \,
    P_\omega(\theta_b, \boldsymbol{\mu}) 
    \left\langle 
    \frac{f\left(\Omega(\theta_b, \beta), \beta, \boldsymbol{\mu}\right)}{|\partial\omega^{(m)}/\partial\Omega|}
    \right\rangle_\beta
    f(\theta_b)\diff{\theta_b}\Diff{3}{\boldsymbol{\mu}} .
    \label{eq: dirac delta on Omega}
\end{equation}
Both approaches have their advantages: Eq.~(\ref{eq: dirac delta on theta}) is more suitable for analytic analysis, since we know the analytic form of $P_\omega$; while Eq.~(\ref{eq: dirac delta on Omega}) has the advantage that we end up with only twofold integrals. We implement {\tt SpyDust} with Eq.~(\ref{eq: dirac delta on Omega}).
More details and the exact form of the integrals are presented in Appendix~\ref{sect: detailed calculations}.

In summary, we adopt a ``hierarchical ensemble averaging'' strategy: first, we average the radiative power over the external alignment characterised by $\theta_L$. This is straightforward for the assumed isotropic external alignment systems.
We also marginalise the grain size in $f(\Omega, a, \beta, \boldsymbol{\mu})$ and then perform the $\beta$ averaging.
Finally, we integrate the product of the emission power and the distribution function over $\theta_b$ and $\boldsymbol{\mu}$.

\begin{figure}
    \centering
    \includegraphics[width=\linewidth]{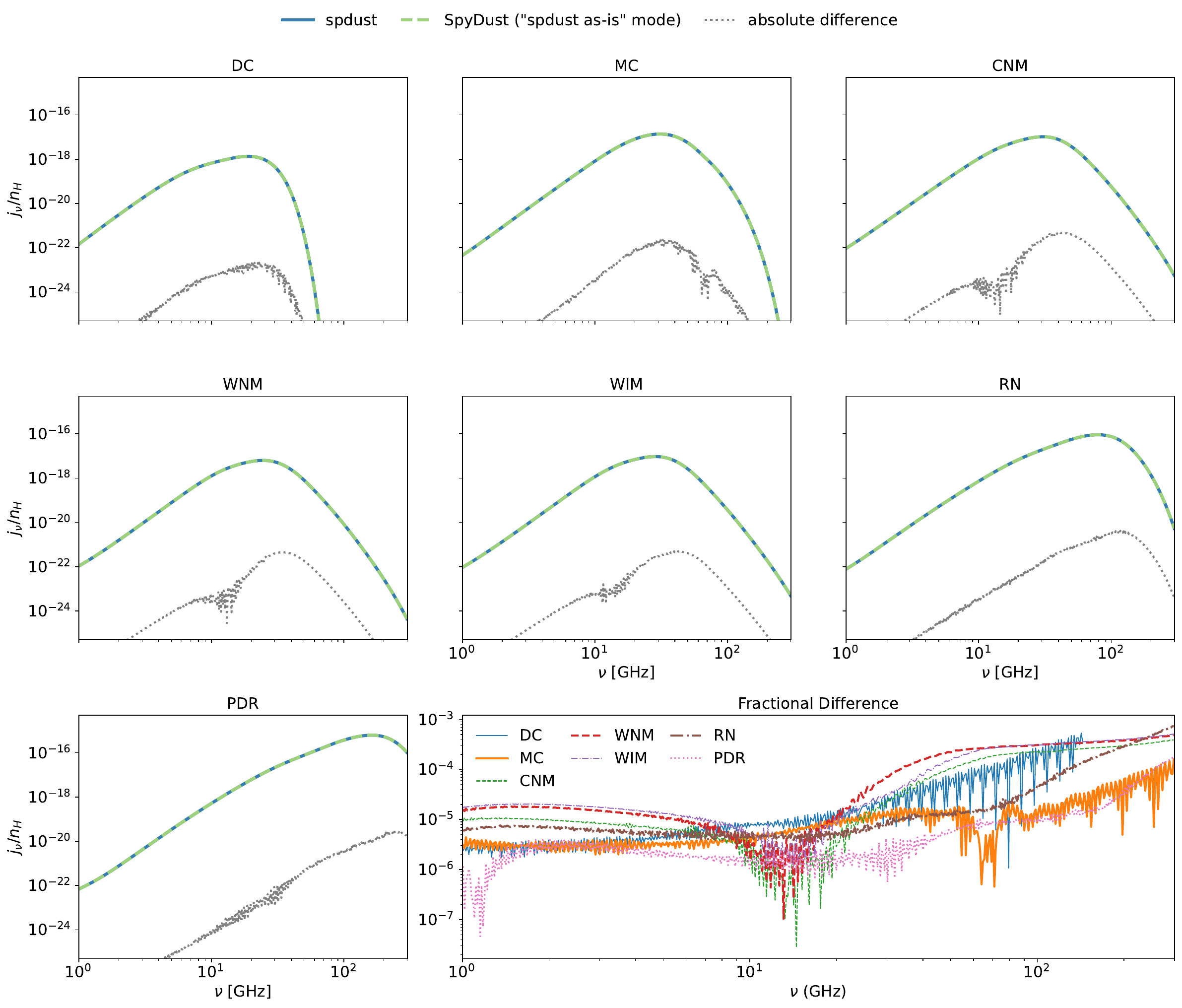}
    \caption{Comparison of the SEDs generated by the ``spdust as-is'' mode in {\tt SpyDust} and the original IDL implementation of spdust. The solid and dashed lines show the SEDs produced by each approach, while the dotted lines show the fractional difference between them as a function of frequency. Overall, the two SEDs show good agreement for all environments, with higher deviations at higher frequencies due to small numerical differences that are multiplicative of the angular momentum powers. These high frequency discrepancies grow modestly with frequency, but remain insignificant as the SED itself decays exponentially at high frequencies. This comparison validates the consistency of the ``spdust as-is'' mode with the original implementation and establishes a baseline for evaluating updates in the {\tt SpyDust}.
    These idealized interstellar environments are defined in \cite{DL98b}, Table~1. 
    }
    \label{fig:spdust as is comparisons}
\end{figure}

\subsection{Comparisons between {\tt SpyDust} and {\tt spdust}}
To accommodate users with different goals, including reproducing the results of this paper, {\tt SpyDust} offers two modes. The first mode, called ``spdust as-is'', provides a direct and as accurate as possible Python translation of the original IDL {\tt spdust}. The second mode, which is the main part, called ``SpyDust'', contains updates and extensions (as discussed in previous sections) for enhanced functionality. In following sections, we systematically compare {\tt SpyDust} with {\tt spdust}.

\subsubsection{Consistency with {\tt spdust}}
First, we evaluate the consistency between the SED generated by our ``spdust as-is'' run and the original IDL results. Figure~\ref{fig:spdust as is comparisons} shows the SEDs from both approaches along with their fractional difference. Overall good agreement is observed. Although the fractional difference tends to increase with frequency - primarily due to the tiny differences in dissipation and fluctuation rates, which scale with powers of frequency\footnote{More specifically, if $f_{1(2)}\sim \exp{(-p_{1(2)}\omega^n)}$, then the fractional difference can be approximated as  $\frac{f_1(\omega)-f_2(\omega)}{f_1(\omega)}\simeq 1-\exp(-\delta p \omega^n)\simeq \delta p \omega^n$. } -
it is ultimately limited by the exponential decay of the spectral energy distribution at high frequencies. As the SED rapidly approaches zero at higher frequencies, these fractional differences less important and negligible. 
Furthermore, the high frequency end is not a primary focus for {\tt SpyDust} because the Fokker-Planck equation does not accurately describe the physics of very small grains (which cause the high frequency emission), which are more susceptible to impulsive torques.
While the fractional difference should not be interpreted as an error in either package\footnote{However, we have fixed two typos in the IDL {\tt spdust} to achieve this level of agreement: (1) In ``plasmadrag.pro'', line $582-583$, ``max(0d, phi\_coeff)''$\rightarrow$``max([0d, phi\_coeff])'' and similarly ``min(1d, phi\_coeff)'' $\rightarrow$ ``min([1d, phi\_coeff])''; (2) In ``charge\_dist.pro'', line $473$, ``W = 4.4'' $\rightarrow$ ``W = 4.4d''.}, understanding this difference quantitatively is valuable. It serves as a baseline to assess whether the corrections and extensions implemented in {\tt SpyDust} introduces trivial or significant changes in the results.

\begin{figure}
    \centering
    \includegraphics[width=0.96\columnwidth]{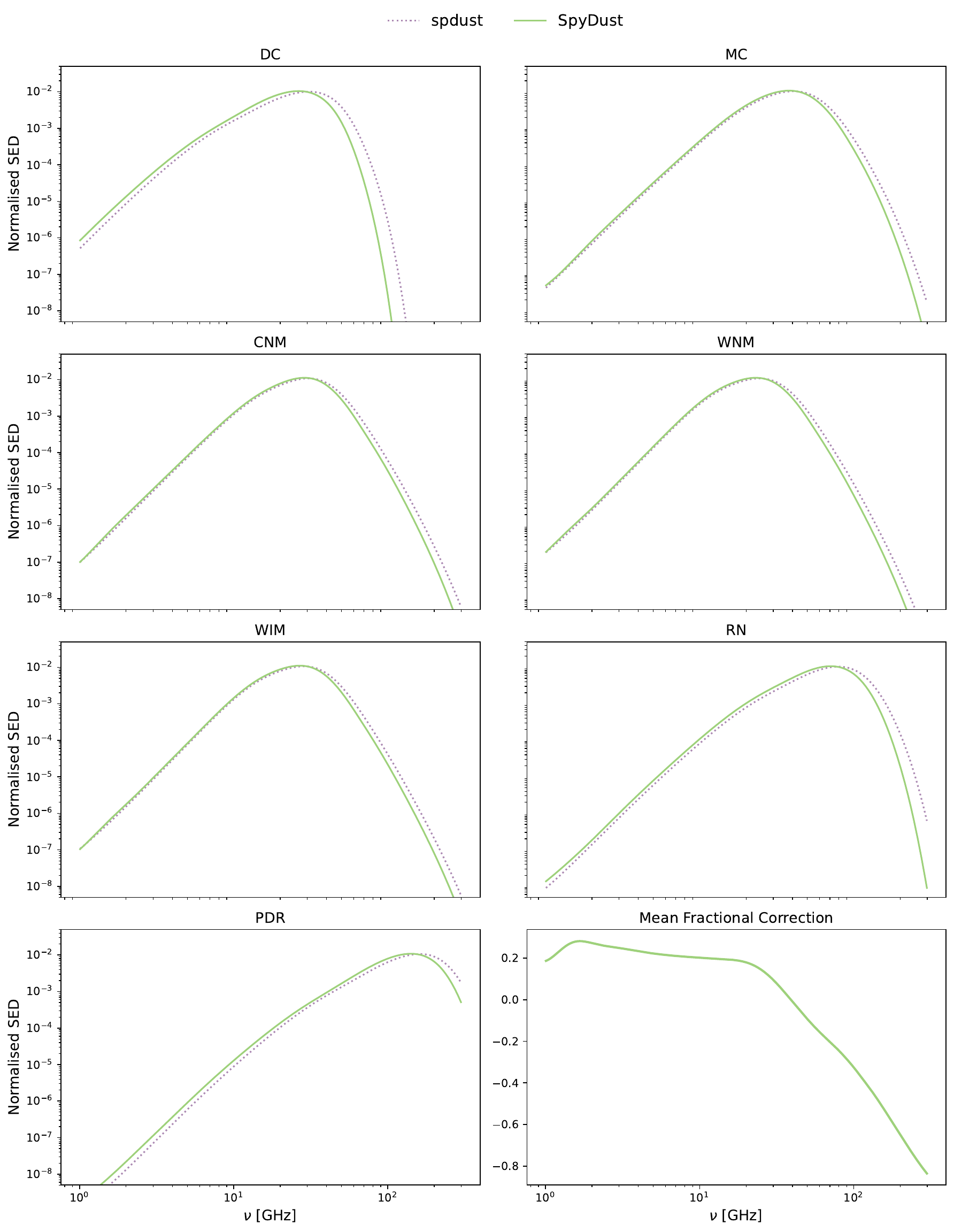}
    \caption{Comparison of SEDs under various correction stages in {\tt SpyDust}. 
    The first seven panels show normalized SEDs for different ISM phases: the original {\tt spdust} model (dotted line) and the updated model with corrections for frequency mapping and electric dipole radiation drift and the $\beta$ generalised plasma drag (solid line). 
    The last panel shows the fractional difference (averaged over environments) between the corrected model and the original model, highlighting non-trivial changes such as a slight leftward shift in peak frequency and modified damping at high frequencies.}
    \label{fig:updated_SED_comparisons}
\end{figure}

\subsubsection{Correction and extension in {\tt SpyDust}}
In this section, we summarize the updates and extensions implemented in {\tt SpyDust}. These modifications can be categorized into three main areas. We will first summarize these modifications, followed by a comparative analysis.
\begin{enumerate}
    \item \textbf{Corrected Mapping between Rotation Frequency and Spectral Frequency}

    While {\tt spdust} models smaller spinning dust grains as discs of finite thickness when considering angular momentum transfer processes, it assumes an idealised case with  $I_1 = I_2 = I_3/2$  (i.e.,  $\beta = -1/2 $, or effectively zero thickness) for emissivity calculations. However, for grains with  $a < 6\text{ \AA}$ and thickness about the typical separation between graphene layers, the $\beta$ actually ranges from $-0.47$ to $-0.39$. This leads to a modified mapping between rotational frequency and spectral frequency, which {\tt SpyDust} corrects.

    \item \textbf{Updated Electric-Dipole Radiation Backreaction and Plasma Drag}
    
    In Section~\ref{sec: radiative damping}, we proposed a revised and generalised formalism for the angular momentum dissipation rate due to electric-dipole radiation backreaction. This includes the correction of a typo in {\tt spdust}, which previously misapplied the tumbling case rate to spherical grains.
    In Section~\ref{sec: plasma}, we updated the $\beta$-generalised expressions for the effects of plasma drag.

    \item \textbf{Extended Shape Distribution for Grains}

    Whereas the original {\tt spdust} assigned a fixed shape (i.e., a single $\beta$ determined by $a$) to each grain size, our formalism allows for a more general shape distribution (an ensemble of $\beta$ values determined by $a$). 

\end{enumerate}
The effect of the first two updates on the SED shape is shown in Figure~\ref{fig:updated_SED_comparisons}. Note that we keep the SED magnitude abstract, i.e. we focus only on the spectral shape. For ease of comparison, we plot the normalised SED. For different ISM phases, we show the original {\tt spdust} SED and the SED after all corrections (including the updated rotation-frequency mapping, the radiative damping and the plasma drag effect).  These updates result in the following changes: the revised mapping and updated drift rate shift the peak frequency slightly to the left, and some phases exhibit faster damping at higher frequencies.  
These changes are not trivial, as can be seen by comparing the average fractional correction in Figure~\ref{fig:updated_SED_comparisons} with the numerical fractional difference in Figure~\ref{fig:spdust as is comparisons}.

\begin{figure}
    \centering
    \includegraphics[width=0.98\textwidth]{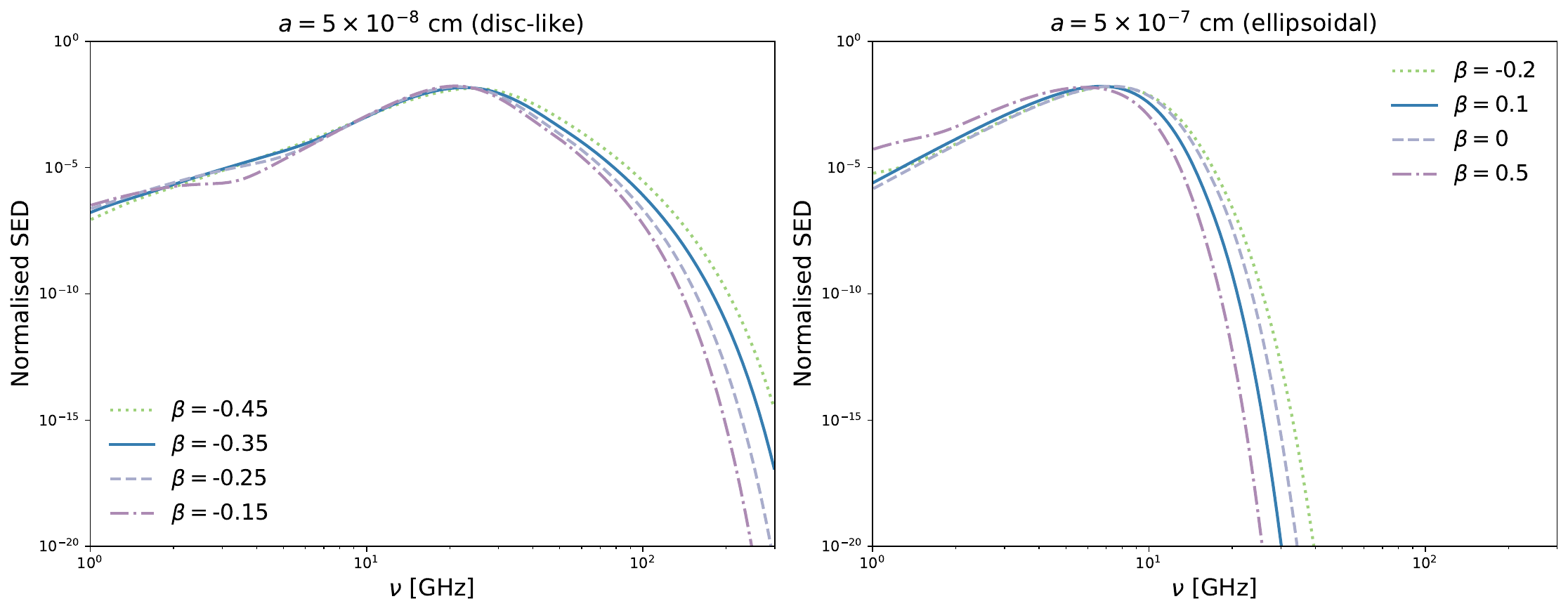}
    \caption{Normalised SED of a single grain with different $\beta$ values (i.e., different grain oblateness).
    For both cylindrical (disc-like) grains and ellipsoidal grains, the oblateness significantly impacts the attenuation of the SED at high frequencies. We use the same CNM environment for all the SEDs.}
    \label{fig:single grain SED different beta}
\end{figure}

\begin{figure}
    \centering
    \includegraphics[width=0.9\columnwidth]{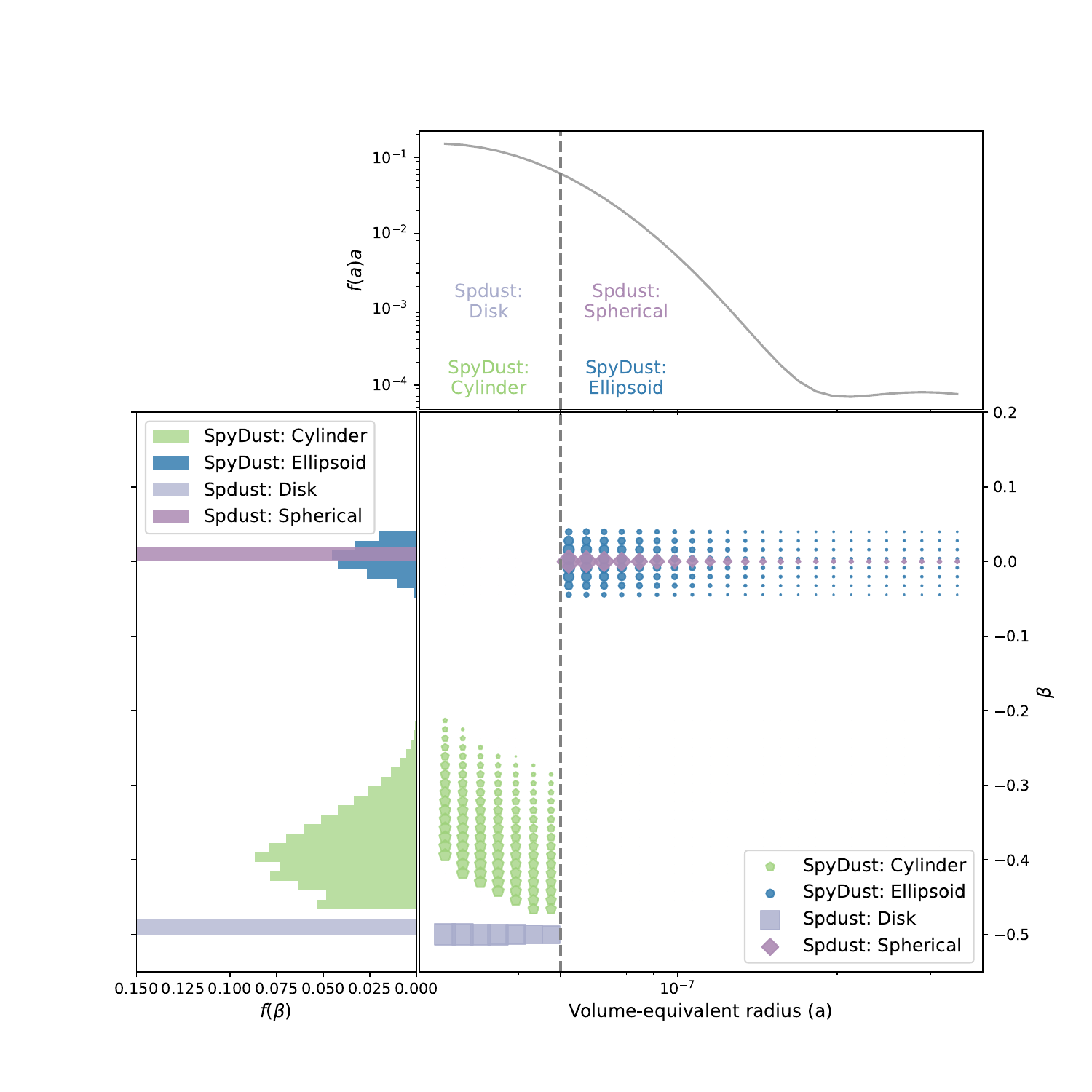}
    \caption{Grain size and shape distributions in {\tt SpyDust} ``$\beta$ ensemble'' model compared with the original ``single-$\beta$'' model from {\tt spdust}.
    In the $\beta$ ensemble, larger grains are represented by a Gaussian distribution centred at $0$ with $\sigma = 0.025$, while small grains is in the right half of the Gaussian distribution centred at $\beta_\text{min}$ (determined by the minimum disc thickness) with $\sigma = 0.1$.
    These toy models are inspired by a ``why not'' approach, with no particular reason it must follow this specific form, and illustrate the new capabilities of {\tt SpyDust}.
    }
    \label{fig: shape size dist}
\end{figure}

\begin{figure}
    \centering
    \includegraphics[width=0.95\linewidth]{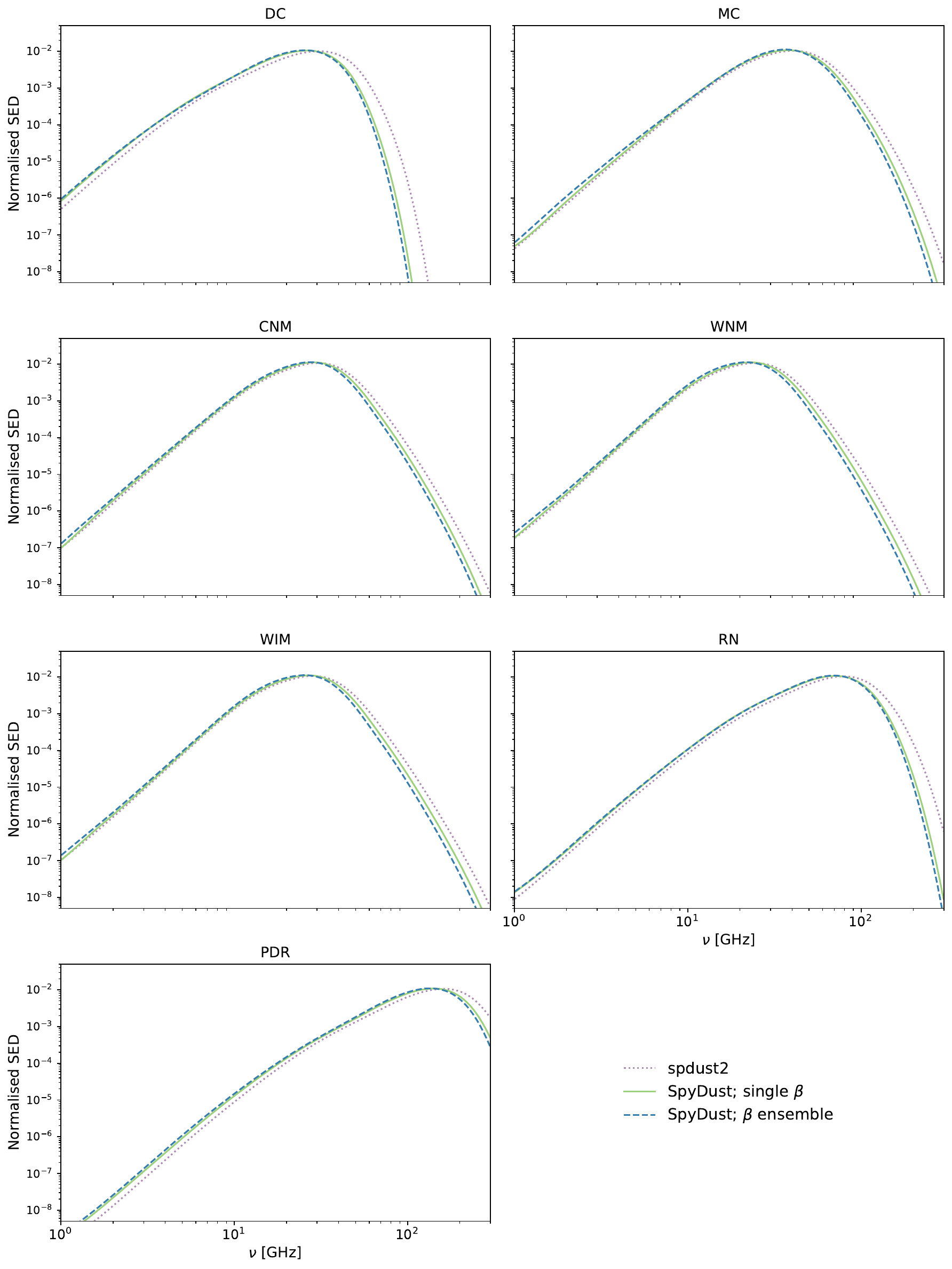}
    \caption{Comparison of SEDs generated using the $\beta$ ensemble model (dashed line) and the single $\beta$ model (solid line for {\tt SpyDust} and dotted line for {\tt spdust}). The $\beta$ ensemble SED shows increased emission at lower frequencies and reduced emission at higher frequencies, reflecting the effect of a wider range of grain shapes on the spectral energy distribution. These idealized interstellar environments defined in \cite{DL98b}, Table~1. }
    \label{fig: beta_SED_comparison}
\end{figure}

The effect of varying grain oblateness, characterized by $\beta$, is illustrated in Figure~\ref{fig:single grain SED different beta}. It is evident that, whether for cylindrical grains (disc-like) or ellipsoidal grains, the oblateness significantly impacts the attenuation of the SED at higher frequencies. For example, the flattest disc-like grains exhibit the broadest SED.
The effect of the possible distribution over grain shape is studied with a toy model, where we extend larger grains from a single spherical shape ($\beta = 0$) to a Gaussian distribution centred at $0$ with $\sigma = 0.025$. 
Smaller grains are assumed to be in the right half of the Gaussian distribution centred at $\beta_\text{min}$ (defined by the minimum disc thickness) with $\sigma = 0.1$. Figure~\ref{fig: shape size dist} shows the grain size and shape distributions in this toy model compared to the original grain size distributions in {\tt spdust}. Referred to as the ``$\beta$ ensemble'' and ``single $\beta$'' respectively, the corresponding SEDs are shown in Figure~\ref{fig: beta_SED_comparison}. We observe that the $\beta$ ensemble affects both the low and high frequency ends of the SED: overall, the $\beta$ ensemble SED shows enhanced low-frequency emission and reduced high-frequency emission compared to the single $\beta$ SED.

\subsection{Degeneracy of parameter space}
\label{sec: parameter space degeneracy}
Although the modelling of spinning dust emission involves more than a dozen parameters, its spectral shape is relatively simple, and even different ISM phases can produce similar spinning dust spectra. Furthermore, as shown in the study of different angular momentum transfer processes, different Maxwellian-type processes can be effectively described by synthesised drift and fluctuation parameters. These indications point to significant degeneracies within the parameter space of spinning dust emission models, opening the possibility for further compression.

\begin{figure}
    \centering
    \includegraphics[width=\linewidth]{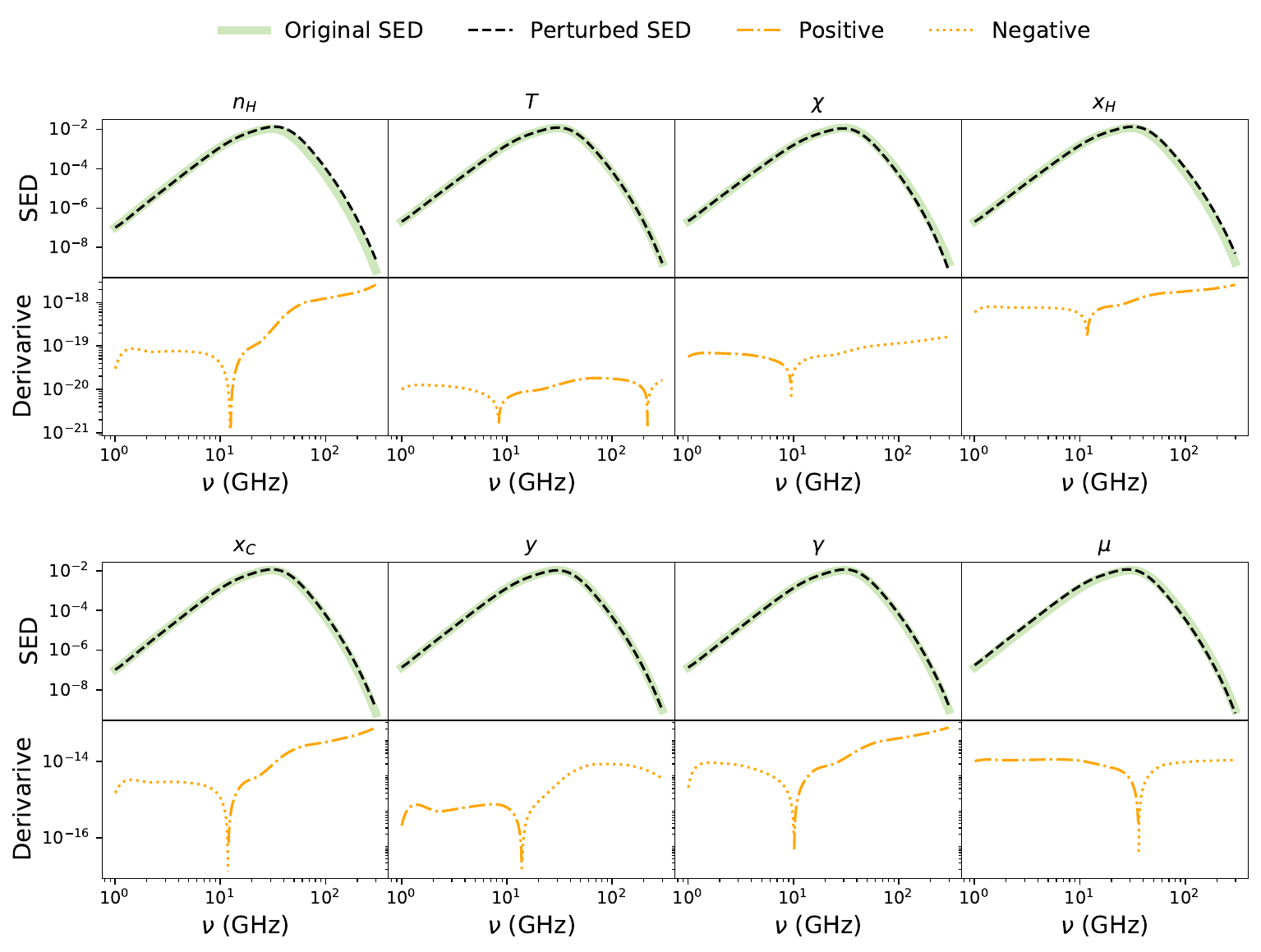}
    \caption{Spectra, perturbed spectra and derivative spectra showing the response of the SED to variations in each parameter within the CNM environment. Each subplot pair shows the perturbed SED and the  derivative of the SED with respect to a parameter.  Dot-dashed lines indicate positive sensitivities, while dotted lines indicate negative sensitivities. These plots show that certain parameters contribute similarly to the shape of the SED, suggesting degeneracies in parameter space (in the perturbation scenario).
    Physical parameters (displayed as `the pivot value $\rightarrow$ the perturbed value'): 
   total hydrogen number density $n_H$~($\text{cm}^3$): $30\rightarrow40$, gas temperature $T$~(K): $100 \rightarrow 150$, ralative radiation field intensity $\chi$: $1 \rightarrow 1.5$, hydrogen ionization fraction $x_H$: $1.2\times 10^{-3} \rightarrow 1.8\times 10^{-3}$, ionized carbon fractional abundance $x_C$: $3\times 10^{-4} \rightarrow 4\times 10^{-4}$, molecular hydrogen fractional abundance $y$: $0 \rightarrow 0.3$, $\text{H}_2$ formation efficiency $\gamma$: $0 \rightarrow 0.3$, rms dipole moment for dust grains $\mu$: $9.3 \rightarrow 10.5$.
    }
    \label{fig:derivative SEDs}
\end{figure}

\begin{figure}
    \centering
    \includegraphics[width=0.7\linewidth]{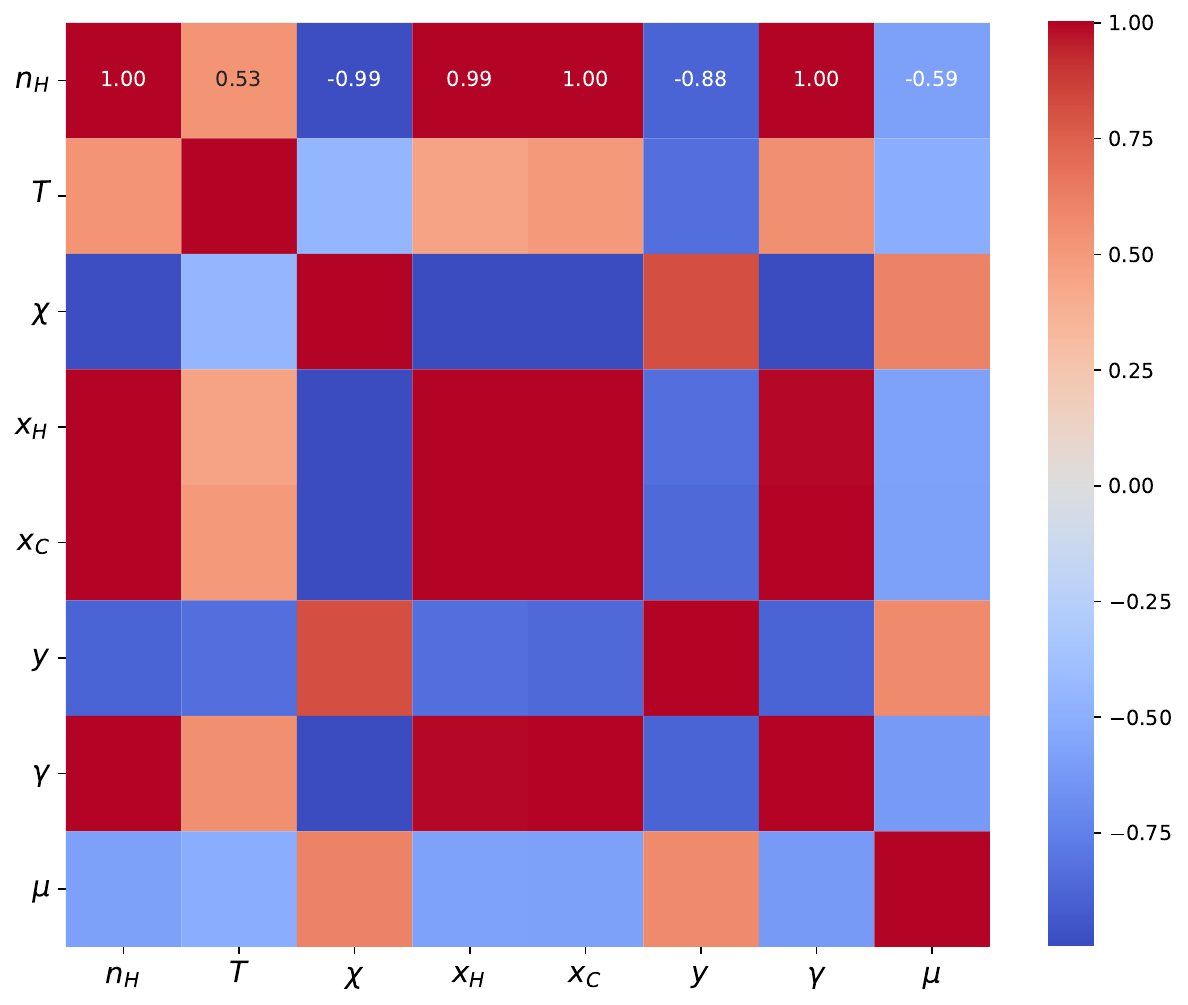}
    \caption{Covariance matrix heatmap showing the correlation structure among parameters in the CNM environment.
    Each cell represents the normalized inner product between the derivative SEDs  for a given parameter pair (see Eq.~(\ref{eq: define covariance})), highlighting the degree of correlation. 
    The high degree of correlation (or anticorrelation) among certain parameters suggests significant degeneracies within the neighborhood parameter space of the given CNM environment. 
    The parameters are as follows: total hydrogen number density $n_H$~($\text{cm}^3$), gas temperature $T$~(K), intensity of the radiation field relative to the average interstellar radiation field $\chi$, hydrogen ionization fraction $x_H$, ionized carbon fractional abundance $x_C$, molecular hydrogen fractional abundance $y$, $\text{H}_2$ formation efficiency $\gamma$, rms dipole moment for dust grains $\mu$.
    }
    \label{fig:cov heatmap}
\end{figure}

\begin{figure}
    \centering
    \includegraphics[width=\linewidth]{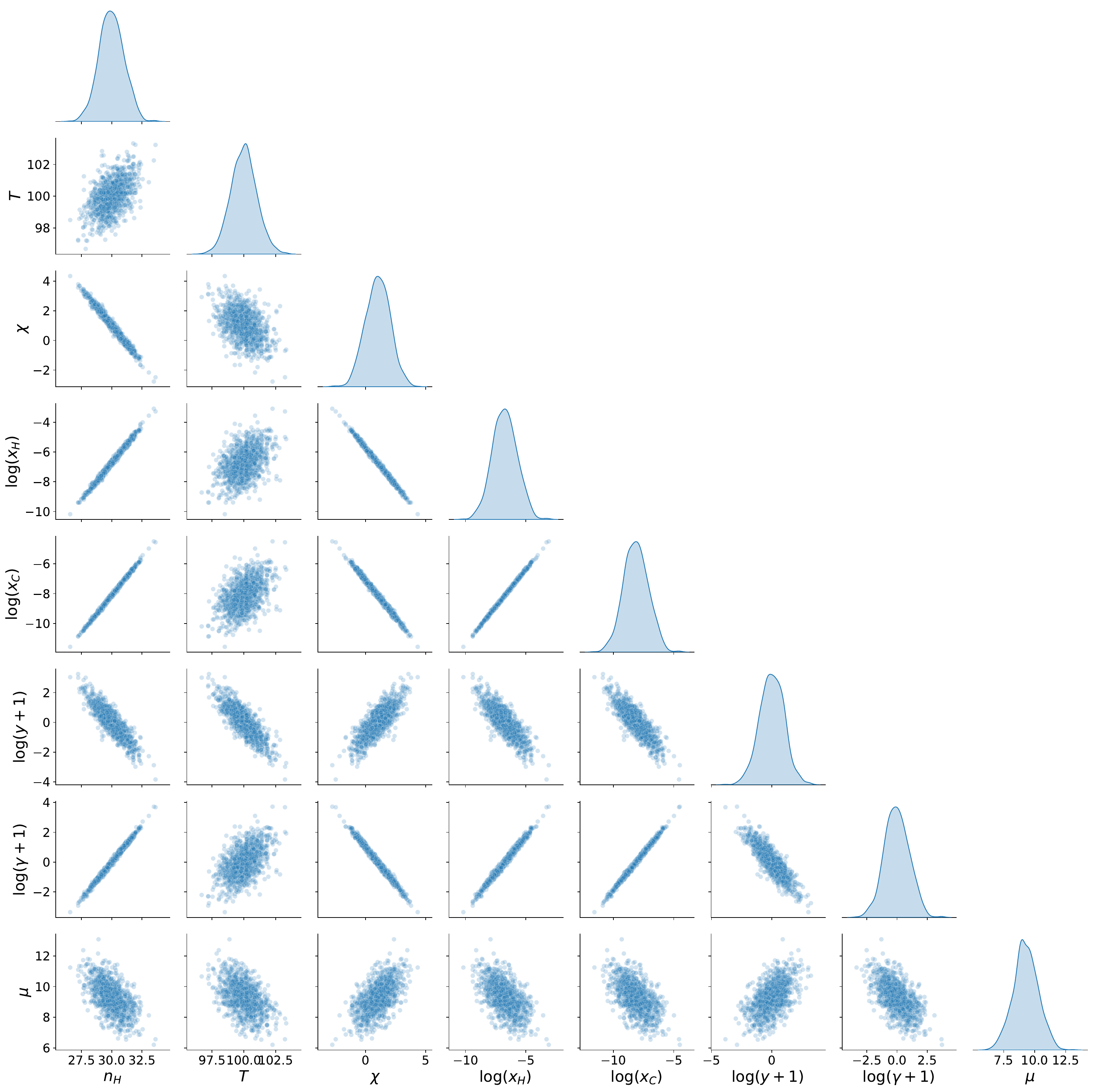}
    \caption{Corner plot showing pairwise relationships between parameters in the neighbourhood parameter space of the given CNM environment, with confidence ellipses indicating the covariance structure for each parameter pair. 
    Diagonal plots show the marginalised distributions for each individual parameter, while off-diagonal plots show the joint distributions, revealing parameter correlations and potential degeneracies. 
    Strong elliptical profiles in certain plots indicate highly correlated parameter pairs, such as the positive correlations of $n_H$ with $x_H$, $x_C$ and $\gamma$, and the negative correlations with $\chi$ and $y$. 
    The parameters are as follows: total hydrogen number density $n_H$~($\text{cm}^3$), gas temperature $T$~(K), intensity of the radiation field relative to the average interstellar radiation field $\chi$, hydrogen ionization fraction $x_H$, ionized carbon fractional abundance $x_C$, molecular hydrogen fractional abundance $y$, $\text{H}_2$ formation efficiency $\gamma$, rms dipole moment for dust grains $\mu$.}
    \label{fig: the corner plot}
\end{figure}

To investigate the degeneracies of these parameters in characterizing the spectral shape, we use the CNM environment as an example. By applying a perturbative approach, we obtain the derivatives of the SED with respect to each parameter, which we refer to as the \textit{derivative SED}. We perturbed all eight parameters except for the grain size parameter, and the resulting derivative SEDs are shown in Figure~\ref{fig:derivative SEDs}.
We define the covariance between parameters using the inner product of normalized derivative SEDs for each parameter pair:
\begin{equation}
    \mathrm{cov}(p_1, p_2) \equiv
    \frac{\langle \partial_{p_1}\ln I_{\nu},\partial_{p_2}\ln I_{\nu}\rangle}{|\partial_{p_1}\ln I_{\nu}||\partial_{p_2}\ln I_{\nu}|}
    \label{eq: define covariance}
\end{equation}
where the inner product $\langle \partial_{p_1} \ln I_{\nu}, \partial_{p_2} \ln I_{\nu} \rangle$ is defined as the integral over frequency (or actually the numerical sum) of the product of the logarithmic SED derivatives with respect to the parameters $p_1$ and $p_2$:
\begin{align}
    \langle \partial_{p_1} \ln I_{\nu}, \partial_{p_2} \ln I_{\nu} \rangle &= \int \frac{\partial\ln I_{\nu}(p_1)}{\partial_{p_1}} \frac{\partial\ln I_{\nu}(p_2)}{\partial_{p_2}} \, d\nu ,
    &
    |\partial_{p_1} \ln I_{\nu}|&\equiv
    \sqrt{\langle \partial_{p_1} \ln I_{\nu}, \partial_{p_1} \ln I_{\nu} \rangle}
    .
\end{align}
This covariance matrix, which captures the parameter dependencies, is shown as a heatmap in Figure~\ref{fig:cov heatmap}. To further illustrate the degeneracies between the parameters, we also present a pair plot with confidence ellipses (see Figure~\ref{fig: the corner plot}), which provides a more intuitive visual summary.
Although this covariance matrix does not correspond to the covariance of Gaussian random variables or the Hessian at the extreme of any likelihood function, it effectively measures the similarity in the response of the SED to each parameter, thereby revealing the underlying degeneracies in parameter space.

\begin{figure}
    \centering
    \includegraphics[width=\linewidth]{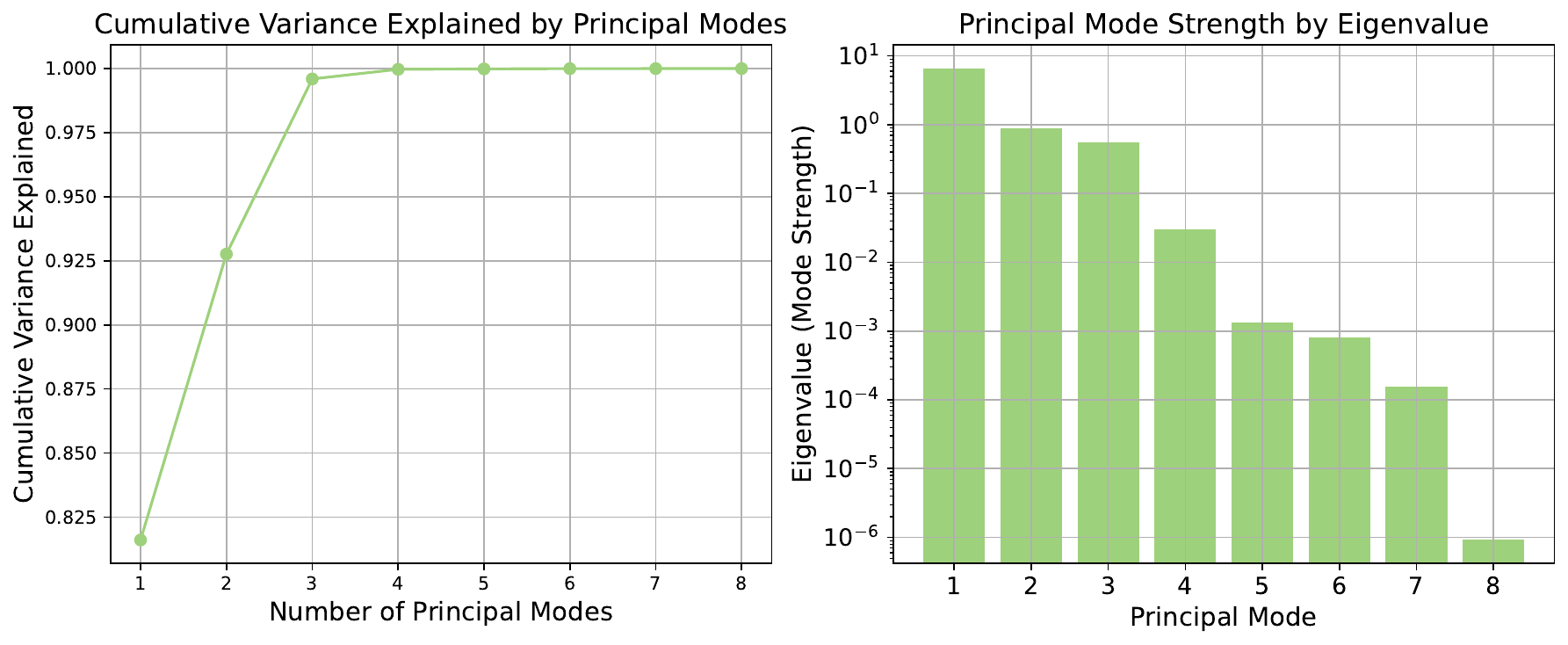}
    \caption{Principal component analysis of the SED structure of the CNM environment. 
    Left: Cumulative variance explained by principal modes. This plot shows the cumulative proportion of total variance captured as each successive principal mode is included. The first four modes together account for 99.8\% of the total variance (on logrithmic scale), highlighting that only a few modes are needed to represent almost all of the variability in the data.
    Right: Principal mode strength by eigenvalue. This bar chart shows the strength of each principal mode represented by its eigenvalue. }
    \label{fig: PCA analysis of perturbed CNM}
\end{figure}

Both the plots of each derivative SED and the covariance matrix clearly reveal significant degeneracies within the parameter space in the neighborhood of the given CNM environment. For example,  $n_H$  shows a strong positive correlation with  $x_H$ ,  $x_C$ , and  $\gamma$ , while it is strongly negatively correlated with  $\chi$  and  $y$.
This suggests that, although the model depends on multiple parameters, the actual number of significant modes in the spectral energy distribution (SED) may be much smaller. To verify this, we conducted a Principal Component Analysis (PCA) on the covariance matrix of these parameters, as shown in Figure~\ref{fig: PCA analysis of perturbed CNM}. In this example, we find that two modes are sufficient to capture nearly all of the variability in the data to a high level of accuracy. We reiterate, however, that this discussion pertains to a perturbed scenario, reflecting the degeneracy within the ``neighbourhood'' parameter space of the given CNM environment. 

Note also that in order to illustrate the SED space degeneracy of the parameters in a simpler way, in this analysis of the environmental parameters we have followed the {\tt spdust} setup by fixing the grain size distribution and the thickness of the smaller grains. However, this does not mean that the size and shape distribution of the grains has no significant effect on the shape of the SED; on the contrary, as clearly shown in Figure~\ref{fig:single grain SED different beta}, the response of the single grain SED to changes in $\beta$ shows a noticeable effect.
More general explorations and applications of moment expansion approaches will be considered in a follow-up paper.

\begin{figure}
    \centering
    \includegraphics[width=\linewidth]{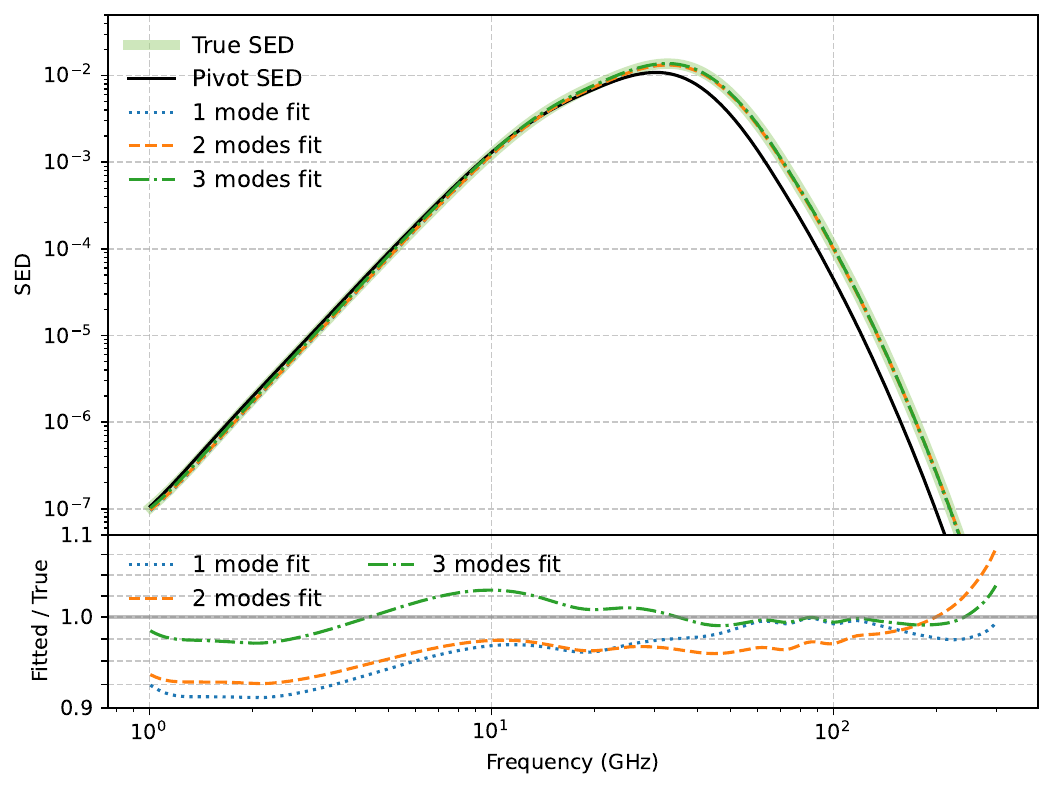}
    \caption{ 
    Fitted the SED using the principal modes derived from principal component analysis [see Eq.(\ref{eq: define covariance}) and Figure\ref{fig:cov heatmap}].
    The top panel displays the true SED (thick line), the pivot SED (thin solid line, representing the SED at the pivot values where the moment expansion is performed), and the fitted SEDs using different numbers of principal modes: the first principal mode (dotted line), the first two principal modes (dashed line), and the first three principal modes (dash-dot line).
    The bottom panel displays the ratio of the fitted SED and the true SED.
    Each principal mode corresponds to a linear combination of the derivative spectra, with weights determined by the eigenvector of the respective mode.}
    \label{fig: fitted SED}
\end{figure}

\section{Conclusion}
\label{sec: conclusion}
In this paper we present {\tt SpyDust}, an improved implementation of the spinning dust emission model based on the Fokker-Planck equation, available as an open source Python package. {\tt SpyDust} is not only a successor to {\tt spdust}, but also incorporates a number of corrections and extensions that notably affect the resulting spectral energy distribution (SED) of spinning dust emissions.

Before summarising these improvements, it is useful to review the basic assumptions of the Fokker-Planck approach (as implemented in both {\tt spdust} and {\tt SpyDust}). 
This model is fundamentally based on the linear regime of angular momentum transport ($\Delta L \ll L$), which allows the fluctuation-dissipation theorem to hold for quasi-equilibrium systems. 
Since this approach integrates fluctuations and dissipations across all processes as synergised fluctuation and dissipation rates, it implicitly assumes that each process remains in a similar quasi-equilibrium state before and after synergy, otherwise it is inconsistent. 
Note that for large angular momentum changes, nonlinear regimes are entered, and the approach fails, such as for impulsive torques on very small grains. 
Recognition of these limitations helps to clarify the conditions under which the Fokker-Planck approach remains valid.


We started by extending the grain shape parameters. Instead of using the conventional volume-equivalent radius ($a$) with fixed shape, we defined grain shapes in terms of their moment of inertia ($I_{\rm ref}$), in-plane ellipticity ($\alpha$) and axial oblateness ($\beta$). Although these three parameters do not fully specify the grain shape (e.g., volume calculations still need to assume a specific geometry, such as a cylinder or ellipsoid), the electric dipole radiation for an individual grain depends only on these three. Together with the rotational configuration parameters, these three shape parameters fully determine the relationship between rotational and spectral frequencies, as well as the directional radiation intensity. 

Unlike {\tt spdust}'s ``perfect disc'' approximation (with $I_1 = I_2 =I_3/2$ or $\beta=-1/2$) for the small grains, {\tt SpyDust} generalizes mapping between the SED and the rotational distribution, taking this shape dependence into account. However, {\tt SpyDust} also makes the fundamental assumption of $\alpha\simeq 0$,  which effectively constrains the grains to negligible wobble in torque-free rotation. 
This is assumed for numerical feasibility, as the torque-free radiation formula is reduced to four normal (oscillating) modes, which is a great simplification. Despite this assumption, {\tt SpyDust} still provides a more general scenario, as it allows discussion of the $\beta$ parameter. On the other hand, to better accommodate the $\alpha\simeq 0$ assumption, we have adapted the definitions of the parameters as in Eq.~(\ref{eq: tuning grain parameter definition}) to extend the validity.

Following the extension of the parameterised grain shapes, we discussed the synthesised SED and its parameter space in Section~\ref{sec: config distribution} and \ref{Sec: SED}.
A hierarchical ensemble averaging approach was used to compute the SED for any distribution function over the parameter space. 
The inclusion of $\beta$ as an additional parameter has also led us to reconsider the angular momentum transport processes of spinning dust grains. 
While most processes are not directly influenced by $\beta$, they can depend on derived quantities such as the volume-equivalent radius $a$ or the effective area radius. {\tt SpyDust} establishes the correspondence between $I_{\rm ref}$, $\alpha$, and $\beta$ and the parameters used in {\tt spdust}, like $a$, under assumed geometries (either elliptical cylinder or ellipsoid, which extend {\tt spdust}'s perfect-disk and spherical shapes, respectively). The calculations of these processes then proceed following the {\tt spdust} treatment. These generalizations ensure that {\tt spdust} is a special case of {\tt SpyDust} in this context, with the exception that we identified minor typos in the IDL code of {\tt spdust}\footnote{These typos are mentioned earlier in this paper.}.
The angular momentum drift rate caused by the electric dipole radiation back-reaction is one process that exhibits a more complex dependence on $\beta$. We derived the radiative damping caused dissipation rate for general $\beta$ (see Section~\ref{sec: radiative damping} and Figure~\ref{fig:tau_ed_inv}).
We also provide the $\beta$-generalised plasma drag effect.
After these corrections and extensions, we observed non-trivial changes in the SED shape (see Figure~\ref{fig:updated_SED_comparisons}).
The above comparisons are made with a fixed grain size distribution, assuming a unique shape for each grain size, as in {\tt spdust}. However, {\tt SpyDust} can also accommodate arbitrary $\beta$ distributions for grains of a given size ``$a$'', leading to further modifications in the SED shape. 
Figure~\ref{fig:single grain SED different beta} shows the effect of varying $\beta$ on the SED shape (especially the width and high frequency attenuation). Figure~\ref{fig: beta_SED_comparison} illustrates the significance of accounting for the breadth of the $\beta$ distribution.

It is worth noting that although these corrections and extensions require more complex numerical steps - such as retaining an interpolation function of the rotational distribution to map between rotational frequency and electromagnetic frequency for different $\beta$ values - the parallelization capabilities offered by {\tt Python} (via {\tt mpi4py}) make these calculations straightforward to implement.

It should be emphasized that the corrections and extensions implemented in {\tt SpyDust} introduce more sophisticated numerical procedures. Specifically, the revised mapping between rotation frequency and spectral frequency now exhibits size-dependent variations for disk-like grains. This necessitates separate computation and interpolation of rotation-frequency mappings for different grain sizes - a fundamental departure from the original {\tt SpDust} implementation where this mapping was assumed to be size-independent.

To address the increased computational demands resulting from this complexity, we have implemented parallel processing capabilities through {\tt Python}'s {\tt mpi4py} framework.
Regarding performance evaluation, direct computational comparisons require careful interpretation due to structural differences between the models. Nevertheless, we provide empirical timing data obtained on standard desktop hardware for reference:
{\tt SpDust} requires approximately $0.75$ seconds to compute 500 frequency points;
Our enhanced {\tt SpyDust} implementation (with all corrections) completes the same calculation in $2.3$ seconds using five MPI subprocesses;
The `spdust\_as\_is' mode in {\tt SpyDust} (without rotational-frequency mapping corrections) achieves a reduced runtime of $0.26$ seconds.


Finally, we found that although the SED shape is influenced by numerous parameters - including eight environmental parameters and several grain size and shape distribution parameters - its structure remains inherently simple, with similar SEDs observed across various ISM phases (e.g., CNM, WNM, WIM). Using the CNM phase as an example, we perturbed environmental parameters to derive ``derivative spectra'' and constructed a covariance matrix from their inner products. Strong correlations emerged (see Figures~\ref{fig:cov heatmap}, \ref{fig: the corner plot}, \ref{fig: PCA analysis of perturbed CNM}), indicating that just three principal modes could capture the majority of SED variations [See the fitted SED in Figure~\ref{fig: fitted SED}]. This high degree of parameter space degeneracy may have significant implications for analyses and model fitting in studies of spinning dust emission and anomalous microwave emissions (AME), particularly in SED fitting of cosmological surveys. We plan to explore these new directions in the future.


\appendix

\section{Emission of Rotating Electric Dipole}
\label{append sec: electrix dipole emission}
The $\boldsymbol{E}(\boldsymbol{x},t)$ field radiated by the moving charge $q(\boldsymbol{x}',t')$ can be expressed by 
\begin{equation}
    \boldsymbol{E}(\boldsymbol{x}, t)
    =
    \frac{q}{c}
    \left[\frac{\boldsymbol{n}\times[(\boldsymbol{n}-\boldsymbol{\beta})\times \Dot{\boldsymbol{\beta}}]}{\kappa^3 R}\right] 
    \label{eq: general radiation Efield}
\end{equation}
where we have only considered the acceleration term as $R\gg 1$ and $\boldsymbol{\beta}= \Dot{x}'/c$ is the velocity of the charge and $\Dot{\boldsymbol{\beta}}$ is the acceleration. 
$R= |\boldsymbol{x}'-\boldsymbol{x}|$ is the distance the radiation has travelled, $\boldsymbol{n}=(\boldsymbol{x}-\boldsymbol{x}')/R$ is the direction, and $\kappa  = 1 -\boldsymbol{n}\cdot\boldsymbol{\beta}$.

We can decompose the velocity vector of the charge into the translational velocity, $\boldsymbol{\beta}_c$, of the centre of mass and the rotational velocity vector, $\boldsymbol{\beta}_r$, of the charge with respect to the centre of mass:
\begin{equation}
    \boldsymbol{\beta}= \boldsymbol{\beta}_c + \boldsymbol{\beta}_r,
\end{equation}
and the acceleration vector is decomposed in a similar way.
For a typical rotating dust grain, we have $\beta_r \ll 1$ and $\Dot{\beta}_r \gg 1$. 
In addition, if we further assume that the translational velocity and acceleration of the grain charges are negligible,\footnote{The high translational acceleration of the charge, or vibration of the chemical band, produces radiation predominantly in the infrared band and is negligible for the band of AME.} 
we have $|\boldsymbol{\beta}|\ll 1 \ll |\Dot{\boldsymbol{\beta}}|$ and then by the Taylor expansion of the radiation field (eq~(\ref{eq: general radiation Efield})) with respect to $\beta$ we obtain the leading order term
    \begin{equation}
    \boldsymbol{E}(\boldsymbol{x}, t)
    =
    \frac{q}{c R}
    \left[\boldsymbol{n}\times(\boldsymbol{n}\times \Dot{\boldsymbol{\beta}})\right],
    \label{eq: leading order contribution}
    \end{equation}
and the first order correction
    \begin{equation}
    \delta\boldsymbol{E}(\boldsymbol{x}, t)
    =
    \frac{q}{c R}
    \left[-\boldsymbol{n}\times(\boldsymbol{\beta}\times \Dot{\boldsymbol{\beta}})
    + 3\,(\boldsymbol{n}\cdot\boldsymbol{\beta})
    \boldsymbol{n}\times(\boldsymbol{n}\times \Dot{\boldsymbol{\beta}})
    \right].
    \end{equation}
In addition to this first-order correction, corrections from the translational velocity distributions of the dust grains can also be considered.
In this paper, we have only considered the leading order contribution.

The sum field of all moving charges, deboted by $P_i$, of the dust grain is given by
\begin{equation}
    \boldsymbol{E}(\boldsymbol{x}, t)
    = \sum_{P_i}
    \frac{q_i}{c R}
    \left[\boldsymbol{n}\times(\boldsymbol{n}\times \Dot{\boldsymbol{\beta}}_i)\right] .
    \label{eq: all charges e field}
\end{equation}
By realizing that
\begin{equation}
    \sum_{P_i} q_i\Dot{\boldsymbol{\beta}}_i  
    =
    \frac{1}{c}
    \sum_{P_i} q_i \Ddot{\boldsymbol{r}}_i  
    =
    \frac{\Ddot{\boldsymbol{\mu}}}{c} 
    ,
\end{equation}
where $\boldsymbol{\mu}$ is the electric dipole moment,
we can rewrite Eq.~(\ref{eq: all charges e field}) as
\begin{equation}
    \boldsymbol{E}(\boldsymbol{x}, t)
    =
    \frac{1}{c^2 R}
    \left[\boldsymbol{n}\times\left(\boldsymbol{n}\times  \Ddot{\boldsymbol{\mu}}\right)\right] ,
\end{equation}
which is proportional to the component of $\Ddot{\boldsymbol{\mu}}$ that is perpendicular to $\boldsymbol{n}$.

Another useful observation is that, under the assumption that all charges rotate at the same angular velocity $\boldsymbol{\omega}$ we have
\begin{equation}
    \sum_{P_i} q_i\Dot{\boldsymbol{\beta}}_i  
    = - \frac{\omega^2}{c} \boldsymbol{\mu}_\perp
\end{equation}
where $\boldsymbol{\mu}_{\perp}$ is the part of $\boldsymbol{\mu}$ that is perpendicular to $\boldsymbol{\omega}$
\begin{equation}
    \boldsymbol{\mu}_{\perp}
    =
    -\hat{\omega}\times\left(\hat{\omega}\times  \boldsymbol{\mu}\right).
\end{equation}
Then the leading order radiation field of the moving dust grain can also be expressed with $\boldsymbol{\mu}$ and $\boldsymbol{\omega}$: 
\begin{equation}
\begin{split}
    \boldsymbol{E}(\boldsymbol{x}, t)
    &=
    - \frac{\omega^2}{c^2 R}
    \left[\boldsymbol{n}\times\left(\boldsymbol{n}\times  \boldsymbol{\mu}_\perp\right)\right] .
\end{split}
\end{equation}

\section{Volume-equivalent Radius as Grain Size Parameter}
\label{sec: volume equivalent radius a}
\begin{figure}[ht]
  \subcaptionbox*{}[.5\linewidth]{%
    \includegraphics[width=\linewidth]{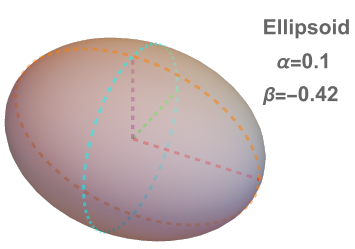}%
  }%
  \hfill
  \subcaptionbox*{}[.5\linewidth]{%
    \includegraphics[width=\linewidth]{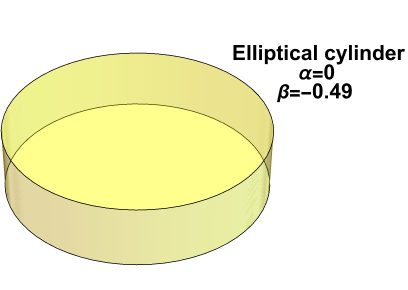}%
  }
  \caption{Examples of the two types of grain.   }
  \label{fig: grain geometry examples}
\end{figure}

In this section we express the volume equivalent radius $a$, the grain size parameter in {\tt spdust}, in terms of the grain parameters $I_{\rm ref}$, $\alpha$ and $\beta$. We consider two geometries: ellipsoidal grains as a generalisation of spherical grains, and elliptical cylinder grains as a generalisation of disc-shaped grains. Figure~\ref{fig: grain geometry examples} shows examples of the two types of grain.

\subsection*{Ellipsoid grain}
For a homogeneous ellipsoidal grain with mass $M$ and principal moments of inertia $I_1, I_2, I_3$, the moments of inertia can be expressed in terms of the radii along the respective axes $r_1, r_2, r_3$ (say $I_1$ for the axis along $r_1$) as follows
\begin{equation}
    I_i = \frac{1}{5} M (r_j^2 + r_k^2)
    \label{eq: ellipsoid grain}
\end{equation}
where $i\neq j\neq k$ and $M=4\pi r_1r_2r_3 \rho /3$.
The mass density of the dust grain is approximately the density of the C atoms, $\rho=\rho_C$, as we consider carbonaceous grains.

The volume-equivalent radius $a$ is defined such that
\begin{equation}
    \frac{4}{3}\pi a^3
    =\frac{M}{\rho}.
\end{equation}
Direct manipulation of eq~(\ref{eq: ellipsoid grain}) yields
\begin{equation}
    a^5 = \frac{15}{8\pi\rho}\left[(I_2 + I_3 - I_1) (I_1 + I_3 - I_2) (I_1 + I_2 - I_3)\right]^{1/3}.
\end{equation}
In the parameterisation with $I_{\rm ref}$, $\alpha$ and $\beta$ and in the limit, $\alpha\approx 0$, we have
\begin{equation}
    a = 
    \left[
    \frac{15 I_{\rm ref}(2\beta+1)^{\frac{1}{3}}}{8\pi\rho(1+\beta)}
    \right]^{\frac{1}{5}}.
\end{equation}

\subsection*{Elliptical cylinder grain}
For a homogeneous elliptical cylinder grain with mass $M$ and principal moments of inertia $I_1, I_2, I_3$, the moments of inertia can be expressed in terms of the radii along the respective axes $r_1, r_2$ and the height $d$ as follows
\begin{align}
    I_1 &=\frac{I_{\rm ref}}{1+\alpha} = \frac{M r_2^2}{4} + \frac{M d^2}{12},
    &
    I_2 &=\frac{I_{\rm ref}}{1-\alpha}= \frac{M r_1^2}{4} + \frac{M d^2}{12},
    &
    I_3 &=\frac{I_{\rm ref}}{1+\beta}= \frac{M(r_1^2 + r_2^2)}{4},
\end{align}
which can be rewritten as
\begin{align}
    d^2 &= \frac{6}{M}(I_1 + I_2 - I_3),
    &
    r_1^2 &= \frac{2}{M}( I_2 + I_3 -I_1 ),
    &
    r_2^2 &= \frac{2}{M}( I_3 + I_1 -I_2 ).
\end{align}
By substituting the above equations with
$\frac{4}{3}\pi a^3  =  \pi r_1 r_2 d = M/\rho$ we obtain
\begin{equation}
    \frac{16}{9} a^6 M^3
    =24
    (I_2 + I_3 - I_1) (I_1 + I_3 - I_2) (I_1 + I_2 - I_3)
\end{equation}
In the parameterization involving $I_{\rm ref}$, $\alpha$ and $\beta$ and in the limit as $\alpha\approx 0$, the expression for $a$ becomes
\begin{equation}
    a 
    = 
    \left[
    \frac{9}{4\pi\rho}
    \frac{I_{\rm ref}}{1+\beta}
    \left(\frac{2\beta+1}{2}\right)^{\frac{1}{3}}
    \right]^{\frac{1}{5}} .
    \label{eq: size parameter of elliptical cylinder grain}
\end{equation}

\section{SED of Each Mode}
\label{sect: detailed calculations}
This Appendix serves as a supplement to Section~\ref{sec: Hierarchical ensemble averaging}, providing the specific forms of the SED for each mode. 
To avoid confusion, we reiterate the marginalization convention used throughout. For example, for a two-dimensional distribution over $a$ and $b$, such as $f(a = a_1) \equiv \int f(a,b)\vert_{a = a_1} \, \mathrm{d}b$.
Additionally, for simplicity, we use $\theta$ without subscripts to represent $\theta_b$.
\paragraph{Mode 1:}
    \begin{itemize}
        \item Mapping between rotation frequency and spectral frequency
        \begin{equation}
            \omega^{(1)}=\Omega.
        \end{equation}
        
        \item Emission averaged over $\theta_L$: 
        \begin{equation}
            \langle P^{(1)} \rangle_{\theta_L}
            =
            \frac{4 }{3 }\omega^4 \mu^2_\parallel  \sin^2\theta .
        \end{equation}
        
        \item Spectral energy density:
        \begin{equation}
        I_\nu^{(1)} = 
        \frac{8 }{9 }\omega^4 
        \int 
        \mu^2_\parallel \, f(\Omega = \omega, \boldsymbol{\mu})
        \Diff{3}{\boldsymbol{\mu}}.
        \end{equation}
        
    \end{itemize}

\paragraph{Mode 2:}
    \begin{itemize}
        \item Mapping between rotation frequency and spectral frequency
        \begin{equation}
            \omega^{(2)}= \Omega
        \left\vert 1+\beta\cos{\theta}\right\vert
        \end{equation}

        \item Emission averaged over $\theta_L$: 
        \begin{equation}
            \langle P^{(2)} \rangle_{\theta_L}
        =
        \frac{1}{3 }\omega^4 \mu^2_\perp  \left(1+\cos{\theta}\right)^2
        \end{equation}

        \item Spectral energy density:
            \begin{equation}
    \begin{split}
    I_\nu^{(2)}
    &= 
    \frac{\omega^4 }{6 } 
    \int \,
    \mu^2_\perp
    \left[
    \int \,
    \delta_D
    \left(
    \omega-\Omega\left\vert 1+\beta\cos{\theta}\right\vert\right)
    f(\Omega, \boldsymbol{\mu}, \beta) 
    \diff{\Omega}
    \right]
    \left(1+\cos{\theta}\right)^2 
    \diff{\cos{\theta}}\Diff{3}{\boldsymbol{\mu}}\diff{\beta} \\
    &= 
    \frac{\omega^4 }{6 } 
    \int \diff{\beta} 
     \diff{\cos{\theta}} \,
    \left[\int
    \mu^2_\perp
    f\left(\Omega = \frac{\omega}{\left\vert 1+\beta\cos{\theta}\right\vert}, \boldsymbol{\mu}, \beta\right)\Diff{3}{\boldsymbol{\mu}}\right]
    \frac{\left(1+\cos{\theta}\right)^2}{\left\vert 1+\beta\cos{\theta}\right\vert} .
    \end{split}
    \end{equation}
    Here we can disregard the singular point at $\beta \cos{\theta} = -1$ because the exponential decay of $f(\Omega\rightarrow\infty)$ effectively suppresses the singular behavior. 

    \end{itemize}

\paragraph{Mode 3:}
    \begin{itemize}
        \item Mapping between rotation frequency and spectral frequency
        \begin{equation}
            \omega^{(3)}=\Omega\left\vert 1-\beta\cos{\theta}\right\vert.
        \end{equation}

        \item Emission averaged over $\theta_L$: 
        \begin{equation}
            \langle P^{(3)} \rangle_{\theta_L}
        =
        \frac{1}{3 }\omega^4 \mu^2_\perp   \left(1-\cos{\theta}\right)^2.
        \end{equation}

        \item Spectral energy density:
         \begin{equation}
    \begin{split}
    I_\nu^{(3)}
    &= 
    \frac{\omega^4 }{6 } 
    \int \,
    \mu^2_\perp
    \left[
    \int \,
    \delta_D
    \left(
    \omega-\Omega\left\vert 1-\beta\cos{\theta}\right\vert\right)
    f(\Omega, \boldsymbol{\mu}, \beta) 
    \diff{\Omega}
    \right]
    \left(1-\cos{\theta}\right)^2 
    \diff{\cos{\theta}}\Diff{3}{\boldsymbol{\mu}}\diff{\beta} \\
    &= 
    \frac{\omega^4 }{6 } 
    \int \diff{\beta} 
     \diff{\cos{\theta}} \,
    \left[\int
    \mu^2_\perp
    f\left(\Omega = \frac{\omega}{\left\vert 1-\beta\cos{\theta}\right\vert}, \boldsymbol{\mu}, \beta\right)\Diff{3}{\boldsymbol{\mu}}\right]
    \frac{\left(1-\cos{\theta}\right)^2}{\left\vert 1-\beta\cos{\theta}\right\vert}. 
    \end{split}
    \end{equation}
    Again, do not worry about the singular point at $\beta \cos{\theta} = 1$.
    \end{itemize}

\paragraph{Mode 4:}
    \begin{itemize}
        \item Mapping between rotation frequency and spectral frequency
        \begin{equation}
            \omega^{(4)}=\Omega
        \left\vert 
        \beta\cos{\theta}\right\vert
        \end{equation}

        \item Emission averaged over $\theta_L$: 
        \begin{equation}
            \langle P^{(4)} \rangle_{\theta_L}
        =
        \frac{2}{3}\omega^4\mu^2_{\perp}   \sin^2 \theta_b .
        \end{equation}

        \item Spectral energy density:
        
    \begin{equation}
    \begin{split}
    I_\nu^{(4)}
    &= \frac{\omega^4}{3 } \int 
    \left[
    \int 
    \,
    \delta_D\left(
    \omega - \Omega
    \left\vert
        \beta\cos{\theta}\right\vert
    \right) 
    \left(
    \int 
    \mu^2_\perp
    f(\Omega, \beta, \boldsymbol{\mu})\Diff{3}{\boldsymbol{\mu}}\right) 
    \diff{\Omega}\right]
    \left(1-\cos^2{\theta}\right) 
    \diff{\cos\theta}\diff{\beta}
    \\
    &= \frac{\omega^4}{3 } \int  \diff{\cos\theta}\diff{\beta}\,
    \frac{1-\cos^2{\theta}}{\left\vert
        \beta\cos{\theta}\right\vert} 
    \int 
    \mu^2_\perp
    f\left(\Omega=\frac{\omega}{\left\vert
        \beta\cos{\theta}\right\vert}, \beta, \boldsymbol{\mu}\right)\Diff{3}{\boldsymbol{\mu}} .
    \end{split}
    \end{equation}

    \end{itemize}
\section{The use of the Fokker-Planck equation}
\label{append: use of FK}
Equation~(\ref{eq: magnitude Fokker Planck}) describes the detailed balance of the $L$-distribution.  
Assuming a homogeneous solution form, one can solve for the distribution function $f(L)$ given the dissipation rate $D_L$ and the fluctuation rate $F_L$.
The general stationary solution is given by
\begin{equation}
        f(L)
        \propto
        \exp{\left[
        \int_0^L \diff{L}\,
        \frac{2D_L - {\partial F_L}/{\partial L}}{F_L}
        \right]}
        \propto
        \frac{1}{F_L(L)}
        \exp{\left[
        \int_0^L \diff{L}\,
        \frac{2D_L}{F_L}
        \right]}.
        \label{eq: general solution fL}
    \end{equation}
When considering angular momentum transfer by several different physical processes, due to the statistical independence of the different angular momentum transport mechanisms, we have to add the dissipation vector and the fluctuation tensor. 

Another use of the Fokker-Planck equation is to solve for the dissipation rate $D_L$ given the stationary distribution $f(L)$ and the fluctuation rate $F_L$. Assuming a homogeneous solution form, the general solution for $D_L$ is given by 
\begin{equation}
    D_L(L)=\frac{1}{2f(L)} \frac{\partial}{\partial L} [F_L(L) f(L)].
    \label{eq: dissipation rate from Fokker-Planck}
\end{equation}
This approach is similar to the detailed balance rule method presented in \cite{SAH11}.

However, regardless of which of the two scenarios is at play, we would like to emphasize the following cautionary points:
\begin{enumerate}
    \item How is $D_L$  derived for each mechanism?

     We can roughly divide $D_L$ definitions into two types: 1. Defined with deterministic physics, where dissipation rates can be rigorously calculated from first principles, such as the torque caused by radiation back-reaction.
	2. Determined with detailed balance, which is usually the case    when more complex random behaviours are involved.
        The distinction between these two types of $D_L$ is clear. We can refer to the former as deterministic $D_L$ and the latter as `$D_L$ in equilibrium with process $X$'.

    \item For $D_L$ defined in equilibrium with process $X$, it is essential to determine the correct equilibrium distribution.

    For spinning dust grains, we must consider, under the assumption of efficient internal thermal fluctuations, the equilibrium between the rotational degrees of freedom and the degrees of freedom of process $X$. 
        
    
    
\end{enumerate}
Following {\tt spdust}, the current plasma treatment first considers the thermal bath only with the plasma heat reservoir, while averaging the rates over the isotropic internal alignment at the end. 
However, the assumed isotropic internal alignment actually means that the internal processes drive the statistical temperature of the $m$ distribution to infinity, while the detailed balance with the plasma heat reservoir alone provides a single temperature for the $\ell$ and $m$ distributions.
A more accurate treatment requires that we explicitly consider the two heat reservoirs with different temperatures: not only the plasma thermal bath, but also the internal thermal bath that drives the $m$-subdistribution to a high temperature.
In a separate paper, we refine the treatment of the plasma drag effect using a detailed Fokker-Planck approximation and explicitly consider both the ionic and internal thermal baths. 

\section{{\tt Spdust} and {\tt SpyDust} Conventions}
\label{app:derivation_Fpl}
In this section, we represent the dissipation rate in {\tt SpyDust} using the notation system of {\tt spdust}. We start by defining the relationship between the rotational distributions in the two different conventions:
\begin{equation}
    f_{\rm {\tt SpyDust}}(L) =
    \frac{4\pi \Tilde{\Omega}^2 }{I_3}
    f_{\tt spdust}(\Tilde{\Omega}),
    \label{eq: convention of dist}
\end{equation}
where $\Tilde{\Omega}=L/I_3$ is the rescaled angular momentum used by {\tt spdust}.
The additional factor is due to the different normalisation conventions between the two frameworks.

The differential equation after the Fokker-Planck approximation in {\tt spdust} is 
\begin{equation}
    \frac{\diff{f_{\tt spdust}}}{\diff{\Tilde{\Omega}}}
    +
    2\frac{ \Tilde{D}}{E_\parallel} f_{\tt spdust}
    =
    0
\end{equation}
where $\Tilde{D}$ and $E_\parallel$ are auxiliary parameters defined with dissipation and fluctuation rates.
The corresponding equation in {\tt SpyDust} is
\begin{equation}
        \frac{\diff
        \left[
        D_L f_{\tt SpyDust}
        \right]}{\diff L}
        =
        \frac{1}{2}
        \frac{\diff^2 
            \left[
            F_L
            f_{\tt SpyDust}
            \right]}{\diff L^2}
        \Rightarrow
        \frac{\diff{f_{\tt SpyDust}}}{\diff{L}}
        +
        \frac{\diff{F_L}/\diff{L} - 2D_L}{F_L} f_{\tt SpyDust}
        =
        0,
        \label{eq: spdust dynamic eq}
\end{equation}
which can be rewritten in terms of $f_{\tt spdust}$ as
\begin{equation}
    \frac{\diff{f_{\tt spdust}}}{\diff{\Tilde{\Omega}}}
    +
    \left[
    \frac{2}{\Tilde{\Omega}}
    +
    \frac{\left(\diff{F_L}/\diff{L} - 2D_L\right)/I_3}{F_L/I_3^2} 
    \right]
    f_{\tt spdust}
    = 0
    \label{eq: SpyDust dynamic eq}
\end{equation}
Comparing Eq.~(\ref{eq: spdust dynamic eq}) and Eq.~(\ref{eq: SpyDust dynamic eq}), we can rewrite $\{\Tilde{D}, E_\parallel\}$ ({\tt spdust} notations) in terms of $\{F_L, D_L\}$ ({\tt SpyDust} notations):
\begin{align}
        E_\parallel &= \frac{F_L}{I_3^2},
        &
        \Tilde{D} &= \frac{ \diff{F_L}/\diff{L} }{2 I_3}
        -\frac{D_L}{I_3}
        + 
        \frac{F_L / I_3^2}{ \Tilde{\Omega}}.
\end{align}
Consequently, we can obtain for {\tt SpyDust} the dimensionless coefficients, $\mathcal{F}$ and $\mathcal{G}$, of the dissipation and fluctuation rates:
\begin{align}
    \mathcal{G}(\Tilde{\Omega})
    &\equiv\frac{I_3\tau_{\rm H}}{2kT}
    \frac{F_L}{I_3^2},
    &
    \mathcal{F}(\Tilde{\Omega})
    &\equiv \frac{\tau_H}{\Tilde{\Omega}} \Tilde{D}.
\end{align}

\section{ Plasma drag and fluctuation}
\label{append: plasma}
In this section we formalise the fluctuation rate due to the plasma effect. We will follow closely the steps in \cite{SAH11}, section~$5.1$, but express the fluctuation explicitly in terms of an arbitrary $\beta$.

The torque exerted on the grain by the passing ions is given by 
\begin{equation}
    \Delta \boldsymbol{L}
    =
    \int_0^{\delta t} \boldsymbol{\mu} \times \boldsymbol{E}\, \diff{t},
\end{equation}
where $\boldsymbol{E}$ is the ambient electric field, which changes rapidly like noise.
Because of the rapid and random nature of $\boldsymbol{E}$, the dissipation rate, $\diff{\langle{\Delta \boldsymbol{L}}\rangle}/\diff{t}$, is difficult to characterise directly. 
In contrast, the fluctuation rate is related to the time-time correlation of $\boldsymbol{E}$, so it can be calculated in terms of the power spectrum of the plasma electric field:
\begin{equation}
    \frac{\diff\langle{\Delta L_i\Delta L_j}\rangle}{\diff{t}} =
    \frac{1}{\delta t}
    \lim_{\delta t\rightarrow 0}
    \int_0^{\delta t}\int_0^{\delta t} \epsilon^{i a b}\epsilon^{j c d} \langle{\mu_a(t)\mu_c(t')}\rangle \langle{E_b(t) E_d(t')}\rangle {\diff{t}\diff{t'}},
\end{equation}
where $a,b,c,d$ (in this section) are abstract spatial component indies, $\epsilon^{ijk}$ denotes the Levi-Civita symbol, and the Einstein summation convention is applied to the repeated indices. We have also taken the approximation that 
\begin{equation}
    \langle{\mu_a(t)\mu_c(t') E_b(t) E_d(t')} \rangle
    \simeq\langle{\mu_a(t)\mu_c(t')} \rangle\langle{E_b(t) E_d(t')}\rangle .
\end{equation}
Assuming that the ambient electric field is isotropic and mixed components are uncorrelated:
\begin{equation}
    \langle{E_b(t) E_d(t')}\rangle = \delta_{bd} C_E(t- t'),
\end{equation}
the rate of fluctuation can be rewritten as
\begin{equation}
    \frac{\diff\langle{\Delta L_i\Delta L_j}\rangle}{\diff{t}} =
    \lim_{\delta t\rightarrow 0} \frac{1}{\delta t}
    \int_0^{\delta t} \diff{t} \int_{-t}^{\delta t-t} \diff{\tau}\,\epsilon^{i a b}\epsilon^{j c b} \langle{\mu_a(t)\mu_c(t+\tau)}\rangle C_E(\tau).
\end{equation}
After direct trigonometric manipulations and reductions as well as $\phi$ and $\psi$ averaging, we obtain 
\begin{equation}
    \langle{\mu_a(t)\mu_c(t+\tau)}\rangle = C^{(ac)}_{\mu}(\tau),
\end{equation}
which in the $O'$ frame (where $\hat{e}_{z'}=\hat{L}$) is given by \cite{SAH11} \footnote{We remind the reader that all angles in this section refer to internal alignment.}
\begin{equation*}
\begin{split}
     C^{(xx)}_{\mu}(\tau)
    &= C^{(yy)}_{\mu}(\tau)\\
    &=
    {\mu_\perp^2}  \left[
    \frac{(1-\cos{\theta})^2}{8}\cos{[(\Dot{\phi}-\Dot{\psi})\tau]}
    + \frac{(1+\cos{\theta})^2}{8}\cos{[(\Dot{\phi}+\Dot{\psi})\tau]}
    \right] 
    + \mu_\parallel^2\frac{\sin^2\theta}{2}  \cos{(\Dot{\phi}\tau)}  ,\\
    C^{(xy)}_{\mu}(\tau)
    &={\mu_\perp^2}  \left[
    \frac{(1-\cos{\theta})^2}{8}\sin{[(\Dot{\phi}-\Dot{\psi})\tau]}
    + \frac{(1+\cos{\theta})^2}{8}\sin{[(\Dot{\phi}+\Dot{\psi})\tau]}
    \right] 
    + \mu_\parallel^2\frac{\sin^2\theta}{2}  \sin{(\Dot{\phi}\tau)}  ,\\
    C^{(zz)}_{\mu}(\tau)
    &=
    \mu_\perp^2\frac{\sin^2{ \theta} }{2}  \cos{(\Dot{\psi}\tau)}  
    + \mu_\parallel^2 \cos^2\theta,
    \qquad
    C^{(xz)}_{\mu}(\tau) =  C^{(yz)}_{\mu}(\tau) = 0 ,
\end{split}
\end{equation*}
and $C^{(ab)}_{\mu}(\tau)=C^{(ba)}_{\mu}(-\tau)$ by definition.
As the ambient $E$ field changes rapidly, $C_E(\tau)$ dampens rapidly as $\tau$ increases from $0$. Therefore, the $\tau$ integral can be effectively expressed as from $-\infty$ to $\infty$:
\begin{equation}
    \frac{\diff\langle{\Delta L_i\Delta L_j}\rangle}{\diff{t}} 
    \simeq
    \lim_{\delta t\rightarrow 0} \frac{1}{\delta t}
    \int_0^{\delta t} \diff{t} 
    \int_{-\infty}^{\infty} \diff{\tau}\,\epsilon^{i a b}\epsilon^{j c b} \,C^{(ac)}_{\mu}(\tau)\, C_E(\tau),
\end{equation}
where the above equation can be rewritten as
\begin{equation}
    \frac{\diff\langle{\Delta L_i\Delta L_j}\rangle}{\diff{t}} 
    \simeq
    \int_{-\infty}^{\infty} \diff{\tau}\,\epsilon^{i a b}\epsilon^{j c b} \,C^{(ac)}_{\mu}(\tau)\, C_E(\tau).
\end{equation}
since the second integral no longer depends on the first integral variable $t$.

Next, we explicitly compute the fluctation rate:
\begin{equation}
\begin{split}
    F^{zz}& = \frac{\diff\langle{\Delta L_z^2}\rangle}{\diff{t}} 
    =
    \int_{-\infty}^{\infty} \diff{\tau} \,\left[C^{(xx)}_{\mu}(\tau)+C^{(yy)}_{\mu}(\tau)\right]\, C_E(\tau) \\
    & =
    {\mu_\perp^2}  \left[
    \frac{(1-\cos{\theta})^2}{4} 
    P_E\left(\frac{\Dot{\phi}-\Dot{\psi}}{2\pi}\right)
    + \frac{(1+\cos{\theta})^2}{4}P_E\left(\frac{\Dot{\phi}+\Dot{\psi}}{2\pi}\right)
    \right] 
    + \mu_\parallel^2 {\sin^2\theta} P_E\left(\frac{\Dot{\phi}}{2\pi}\right),
\end{split}
\end{equation}
where $P_E$ is the plasma electric field power spectrum defined by
\begin{equation}
    P_E\left(\nu\right)\equiv 
    \int_{-\infty}^{\infty} \diff{\tau}\, \cos{(2\pi\nu\tau)} C_E(\tau) .
\end{equation}
Similarly, we find
\begin{equation}
\begin{split}
    F^{xx} &= F^{yy} =\frac{\diff\langle{\Delta L_y^2}\rangle}{\diff{t}} 
    =
    \int_{-\infty}^{\infty} \diff{\tau} \,\left[C^{(yy)}_{\mu}(\tau)+C^{(zz)}_{\mu}(\tau)\right]\, C_E(\tau) \\
    &= \frac{F^{zz}}{2} + \mu_\perp^2\frac{\sin^2{ \theta} }{2}  
     P_E\left(\frac{\Dot{\psi}}{2\pi}\right)  
    + \mu_\parallel^2 \cos^2\theta  \, P_E\left(0\right) .
\end{split}        
\end{equation}
Substituting the frequencies as
\begin{align}
    \frac{\Dot{\phi}}{2\pi}
    &=\nu_{\rm ref}, 
    &
    \frac{\Dot{\psi}}{2\pi}
    &=\nu_{\rm ref}\beta\cos{\theta}, 
\end{align}
where $\nu_{\rm ref} = L/ (2\pi I_{\rm ref})$,
$F^{zz}$ can be rewritten as
\begin{multline*}
    F^{zz}(\boldsymbol{L}, \theta)
     =
    {\mu_\perp^2}  \left[
    \frac{(1-\cos{\theta})^2}{4} 
    P_E\left(\nu_{\rm ref} (1-\beta\cos{\theta})\right)
    + \frac{(1+\cos{\theta})^2}{4}P_E\left(\nu_{\rm ref} (1+\beta\cos{\theta})\right)
    \right] \\
    + \mu_\parallel^2 {\sin^2\theta} P_E\left(\nu_{\rm ref}\right),
\end{multline*}
and the $xx$ and $yy$ components are
\begin{equation}
    F^{xx} = F^{yy} 
    = \frac{F^{zz}}{2} + \mu_\perp^2\frac{\sin^2{ \theta} }{2}  
     P_E\left(\beta\cos{\theta}\nu_{\rm ref}\right)  
    + \mu_\parallel^2 \cos^2\theta  \, P_E\left(0\right) .\end{equation}
The $\theta$-average of the fluctuation rate of the magnitude is thus given by
\begin{equation}
    F_L = \Bar{F}^{zz}
    =
    \frac{\mu_\perp^2}{2}
    \int_{-1}^1(1-x)^2 
    P_E\left(\nu_{\rm ref} (1-\beta x)\right) \diff{x}
     \\
    + \frac{2}{3}\,\mu_\parallel^2 P_E\left(\nu_{\rm ref}\right),
    \label{eq: fluctuation plasma L}
\end{equation}
where we substituted $x=\cos\theta$.

Similarly, the cross-fluctuation between the angular momentum amplitude and its projection onto the grain axis, $\hat{z}_b$, is expressed as:
\begin{equation}
    F^{zz_b}
    \equiv
    \frac{\diff{\braket{\Delta L \, \Delta L_{z_b}}}}{\diff{t}}
    =
    {\mu_\perp^2}  \left[
    \frac{(1+\cos{\theta})^2}{4}P_E\left(\nu_{\rm ref} (1+\beta\cos{\theta})\right)
    -\frac{(1-\cos{\theta})^2}{4} 
    P_E\left(\nu_{\rm ref} (1-\beta\cos{\theta})\right)
    \right] 
\end{equation}
with $\Delta L_{z_b} = \Delta L\cos{\theta}$.
The fluctuation rates driven by the stochastic ambient plasma electric field are formalised as a $\beta$-generalised extension of the calculations in \cite{SAH11}. These rates are then used to derive the angular momentum drift rate. 


\acknowledgments

We thank Yacine Ali-Ha\"{i}moud, 
Roke Cepeda Arroita, Martin Bucher, Phil Bull, Clive Dickinson, Jose Alberto Rubino-Martin and Robert Watson for their positive comments and encouragement.
The results were obtained as part of a project that has received funding from the European Research Council (ERC) under the European Union's Horizon 2020 research and innovation programme (Grant agreement No. 948764; ZZ).  
We furthermore acknowledge support by the RadioForegroundsPlus Project HORIZON-CL4-2023-SPACE-01, GA 101135036.

\bibliographystyle{JHEP}

\bibliography{spydust.bib} 

\providecommand{\href}[2]{#2}\begingroup\raggedright\begin{thebibliography}{10}

\bibitem{erickson1957}
W.C.~Erickson, \emph{A mechanism of non-thermal radio-noise origin.}, {\emph{Astrophysical Journal, vol. 126, p. 480} {\bfseries 126} (1957) 480}.

\bibitem{hoyle1970}
F.~Hoyle and N.C.~Wickramasinghe, \emph{Dust in supernova explosions}, {\emph{Nature} {\bfseries 226} (1970) 62}.

\bibitem{ferrara1994}
A.~Ferrara and R.-J.~Dettmar, \emph{Radio-emitting dust in the free electron layer of spiral galaxies: Testing the disk/halo interface}, {\emph{Astrophysical Journal, Part 1 (ISSN 0004-637X), vol. 427, no. 1, p. 155-159} {\bfseries 427} (1994) 155}.

\bibitem{kogut1995high}
A.~Kogut, A.~Banday, C.~Bennett, K.~Gorski, G.~Hinshaw and W.~Reach, \emph{High-latitude galactic emission in the cobe dmr two-year sky maps}, {\emph{arXiv preprint astro-ph/9509151} (1995) }.

\bibitem{kogut1996microwave}
A.~Kogut, A.J.~Banday, C.L.~Bennett, K.M.~G{\'o}rski, G.~Hinshaw, G.F.~Smoot et~al., \emph{Microwave emission at high galactic latitudes in the four-year dmr sky maps}, {\emph{The Astrophysical Journal} {\bfseries 464} (1996) L5}.

\bibitem{leitch1997anomalous}
E.~Leitch, A.~Readhead, T.~Pearson and S.~Myers, \emph{An anomalous component of galactic emission}, {\emph{The Astrophysical Journal} {\bfseries 486} (1997) L23}.

\bibitem{DL98a}
B.~Draine and A.~Lazarian, \emph{Diffuse galactic emission from spinning dust grains}, {\emph{The Astrophysical Journal} {\bfseries 494} (1998) L19}.

\bibitem{DL98b}
B.~Draine and A.~Lazarian, \emph{Electric dipole radiation from spinning dust grains}, {\emph{The Astrophysical Journal} {\bfseries 508} (1998) 157}.

\bibitem{lazarian2003microwave}
A.~Lazarian and D.~Finkbeiner, \emph{Microwave emission from aligned dust}, {\emph{New Astronomy Reviews} {\bfseries 47} (2003) 1107}.

\bibitem{AHD09}
Y.~Ali-Ha{\"\i}moud, C.M.~Hirata and C.~Dickinson, \emph{A refined model for spinning dust radiation}, {\emph{Monthly Notices of the Royal Astronomical Society} {\bfseries 395} (2009) 1055}.

\bibitem{hoang2010}
T.~Hoang, B.~Draine and A.~Lazarian, \emph{Improving the model of emission from spinning dust: effects of grain wobbling and transient spin-up}, {\emph{The Astrophysical Journal} {\bfseries 715} (2010) 1462}.

\bibitem{SAH11}
K.~Silsbee, Y.~Ali-Ha{\"\i}moud and C.M.~Hirata, \emph{Spinning dust emission: the effect of rotation around a non-principal axis}, {\emph{Monthly Notices of the Royal Astronomical Society} {\bfseries 411} (2011) 2750}.

\bibitem{hoang2011}
T.~Hoang, A.~Lazarian and B.T.~Draine, \emph{Spinning dust emission: effects of irregular grain shape, transient heating, and comparison with wilkinson microwave anisotropy probe results}, {\emph{The Astrophysical Journal} {\bfseries 741} (2011) 87}.

\bibitem{draine1999magnetic}
B.~Draine and A.~Lazarian, \emph{Magnetic dipole microwave emission from dust grains}, {\emph{The Astrophysical Journal} {\bfseries 512} (1999) 740}.

\bibitem{hoang2016unified}
T.~Hoang and A.~Lazarian, \emph{A unified model of grain alignment: radiative alignment of interstellar grains with magnetic inclusions}, {\emph{The Astrophysical Journal} {\bfseries 831} (2016) 159}.

\bibitem{hensley2017modeling}
B.S.~Hensley and B.T.~Draine, \emph{Modeling the anomalous microwave emission with spinning nanoparticles: no pahs required}, {\emph{The Astrophysical Journal} {\bfseries 836} (2017) 179}.

\bibitem{draine2016quantum}
B.~Draine and B.S.~Hensley, \emph{Quantum suppression of alignment in ultrasmall grains: microwave emission from spinning dust will be negligibly polarized}, {\emph{The Astrophysical Journal} {\bfseries 831} (2016) 59}.

\bibitem{dickinson2018state}
C.~Dickinson, Y.~Ali-Ha{\"\i}moud, A.~Barr, E.~Battistelli, A.~Bell, L.~Bernstein et~al., \emph{The state-of-play of anomalous microwave emission (ame) research}, {\emph{New Astronomy Reviews} {\bfseries 80} (2018) 1}.

\bibitem{galloway2023beyondplanck}
M.~Galloway, K.J.~Andersen, R.~Aurlien, R.~Banerji, M.~Bersanelli, S.~Bertocco et~al., \emph{Beyondplanck-iii. commander3}, {\emph{Astronomy \& Astrophysics} {\bfseries 675} (2023) A3}.

\bibitem{kogut2019cmb}
A.~Kogut, M.~Abitbol, J.~Chluba, J.~Delabrouille, D.~Fixsen, J.~Hill et~al., \emph{Cmb spectral distortions: status and prospects}, .

\bibitem{Chluba2021Voyage}
J.~{Chluba}, M.H.~{Abitbol}, N.~{Aghanim}, Y.~{Ali-Ha{\"\i}moud}, M.~{Alvarez}, K.~{Basu} et~al., \emph{{New horizons in cosmology with spectral distortions of the cosmic microwave background}}, \href{https://doi.org/10.1007/s10686-021-09729-5}{\emph{Experimental Astronomy} {\bfseries 51} (2021) 1515} [\href{https://arxiv.org/abs/1909.01593}{{\ttfamily 1909.01593}}].

\bibitem{abitbol_pixie}
M.H.~{Abitbol}, J.~{Chluba}, J.C.~{Hill} and B.R.~{Johnson}, \emph{{Prospects for Measuring Cosmic Microwave Background Spectral Distortions in the Presence of Foregrounds}}, \href{https://doi.org/10.1093/mnras/stw030}{\emph{Monthly Notices of the Royal Astronomical Society} (2017) } [\href{https://arxiv.org/abs/1705.01534}{{\ttfamily 1705.01534}}].

\bibitem{Jens17}
J.~Chluba, J.C.~Hill and M.H.~Abitbol, \emph{{Rethinking CMB foregrounds: systematic extension of foreground parametrizations}}, \href{https://doi.org/10.1093/mnras/stx1982}{\emph{Monthly Notices of the Royal Astronomical Society} {\bfseries 472} (2017) 1195} [\href{https://arxiv.org/abs/https://academic.oup.com/mnras/article-pdf/472/1/1195/19732644/stx1982.pdf}{{\ttfamily https://academic.oup.com/mnras/article-pdf/472/1/1195/19732644/stx1982.pdf}}].

\bibitem{vacher2023high}
L.~Vacher, J.~Chluba, J.~Aumont, A.~Rotti and L.~Montier, \emph{High precision modeling of polarized signals: Moment expansion method generalized to spin-2 fields}, {\emph{Astronomy \& Astrophysics} {\bfseries 669} (2023) A5}.

\bibitem{carones2024optimization}
A.~Carones and M.~Remazeilles, \emph{Optimization of foreground moment deprojection for semi-blind cmb polarization reconstruction}, {\emph{Journal of Cosmology and Astroparticle Physics} {\bfseries 2024} (2024) 018}.

\bibitem{rot_mechanics}
D.~Garanin, ``Rotational motion of rigid bodies.'' \url{https://www.lehman.edu/faculty/dgaranin/Mechanics/Mechanis_of_rigid_bodies.pdf}.

\bibitem{barnett1935gyromagnetic}
S.J.~Barnett, \emph{Gyromagnetic and electron-inertia effects}, {\emph{Reviews of Modern Physics} {\bfseries 7} (1935) 129}.

\bibitem{ali13review}
Y.~Ali-Ha{\"\i}moud et~al., \emph{Spinning dust radiation: a review of the theory}, {\emph{Advances in Astronomy} {\bfseries 2013} (2013) }.

\end{thebibliography}\endgroup

\end{document}